\documentclass
[amsfonts,amssymb,oneside,10pt,a4paper,preprint,pra,onecolumn,nobalancelastpage,nofootinbib,noshowpacs,notitlepage]{revtex4}%
\usepackage{amsfonts}
\usepackage{amsmath}
\usepackage{amssymb}
\usepackage{graphicx}
\usepackage[nonscript]{accents}%
\providecommand{\U}[1]{\protect\rule{.1in}{.1in}}

\ifx\pdfoutput\relax\let\pdfoutput=\undefined\fi
\newcount\msipdfoutput
\ifx\pdfoutput\undefined\else
\ifcase\pdfoutput\else
\ifx\paperwidth\undefined\else
\ifdim\paperheight=0pt\relax\else\pdfpageheight\paperheight\fi
\ifdim\paperwidth=0pt\relax\else\pdfpagewidth\paperwidth\fi
\fi\fi\fi
\begin{document}
\title{Open systems dynamics: Simulating master equations in the computer}
\author{Carlos Navarrete-Benlloch}
\thanks{These notes are a work in progress, so use them with caution. They have been greatly benefited from discussions with Juan Jos\'{e} Garc\'{\i}a-Ripoll and
Diego Porras. Thanks also to Alejandro Gonz\'{a}lez-Tudela for proof-reading and useful suggestions, and to Eli\v{s}ka Greplov\'a for comments, motivation, and putting the code into practice.
\\
E-mail: carlos.navarrete@mpq.mpg.de
\\
Webpage: www.carlosnb.com}

\begin{abstract}
Master equations are probably the most fundamental equations for anyone
working in quantum optics in the presence of dissipation. In this context it
is then incredibly useful to have efficient ways of coding and simulating such
equations in the computer, and in this notes I try to introduce in a
comprehensive way how do I do so, focusing on Matlab, but making it general
enough so that it can be directly translated to any other language or software
of choice. I inherited most of my methods from Juan Jos\'{e}
Garc\'{\i}a-Ripoll (whose numerical abilities I cannot praise enough),
changing them here and there to accommodate them to the way my (fairly
limited) numerical brain works, and to connect them as much as possible to how
I understand the theory behind them. At present, the notes focus on how to
code master equations and find their steady state, but I hope soon I will be
able to update them with time evolution methods, including how to deal with
time-dependent master equations. During the last 4 years I've tested these
methods in various different contexts, including circuit quantum
electrodynamics, the laser problem, optical parametric oscillators, and
optomechanical systems. Comments and (constructive) criticism are greatly
welcome, and will be properly credited and acknowledged.

\end{abstract}
\maketitle
\tableofcontents

\newpage

\section{On the structure of master equations and steady states}

Let me start by briefly introducing in a greatly simplified manner the
concepts of \textit{master equation} and \textit{steady states}. Consider a
system endorsed with a Hilbert space $\mathcal{H}$ of dimension $d$ (since we
are interested in doing numerics, we will always assume that $d$ is finite,
what might need truncating the Hilbert space dimension when this is infinite
in reality). We say that the system is open when it is part of a larger space
with which it exchanges energy, information, etc..., and generically we call
\textit{environment} to the rest of this larger space. In many situations,
most notably when the environment is much larger or evolves much faster than
the system, it is possible to describe the dynamics of the latter via a linear
differential equation for its individual state, which of course is generally
mixed since eliminating the environment means loosing information, hence
requiring a description in terms of a density operator $\hat{\rho}$. We call
master equation to the evolution equation for the system's density operator,
and in the following we will be guided by the generic form\footnote{We will
consider a single jump operator for notational simplicity, but everything
we'll do is generalized straightforwardly to the general irresversible term
$\sum_{j}\Gamma_{j}(2\hat{J}_{j}\hat{\rho}\hat{J}_{j}^{\dagger}-\hat{J}%
_{j}^{\dagger}\hat{J}_{j}\hat{\rho}-\hat{\rho}\hat{J}_{j}^{\dagger}\hat{J}%
_{j})$, or even to terms of different form.}%
\begin{equation}
\frac{d\hat{\rho}}{dt}=-\mathrm{i}\left[  \hat{H},\hat{\rho}\right]
+\Gamma(2\hat{J}\hat{\rho}\hat{J}^{\dagger}-\hat{J}^{\dagger}\hat{J}\hat{\rho
}-\hat{\rho}\hat{J}^{\dagger}\hat{J})\equiv\mathcal{L[}\hat{\rho}],
\label{MasterEq}%
\end{equation}
where $\hat{H}$ is an Hermitian operator containing the system Hamiltonian and
coherent or reversible exchange processes with the environment, while $\hat
{J}$ is a so-called jump operator (with associated rate $\Gamma>0$) which
describes irreversible processes such as excitations which are lost to the
large environment never to come back to the system. $\mathcal{L}$ is then an
linear map usually called \textit{Liouvillian superoperator}, whose name comes
from the fact that it acts on operators to give operators.

Consider a basis $\{|n\rangle\}_{n=1,2,...,d}$, which allows us to represent
the density operator as%
\begin{equation}
\hat{\rho}=\sum_{nm=1}^{d}\rho_{nm}|n\rangle\langle m|,
\end{equation}
with $\rho_{nm}=\langle n|\hat{\rho}|m\rangle$. The master equation, once
projected into this basis, just provides a linear system of ordinary
differential equations for the components of the density matrix, that is%
\begin{equation}
\frac{d\rho_{nm}}{dt}=\sum_{k=1}^{d}\left(  \mathrm{i}\rho_{nk}H_{km}%
-\mathrm{i}H_{nk}\rho_{km}\right)  +\sum_{kl=1}^{d}\left(  2\Gamma J_{nk}%
\rho_{kl}J_{ml}^{\ast}-\Gamma J_{kn}^{\ast}J_{kl}\rho_{lm}-\Gamma\rho
_{nk}J_{lk}^{\ast}J_{lm}\right)  =\sum_{kl=1}^{d}L_{nm;kl}\rho_{kl},
\end{equation}
with%
\[
L_{nm;kl}=-\mathrm{i}H_{nk}\delta_{ml}+\mathrm{i}H_{lm}\delta_{kn}+2\Gamma
J_{nk}J_{ml}^{\ast}-\Gamma(\hat{J}^{\dagger}\hat{J})_{nk}\delta_{ml}%
-\Gamma\delta_{kn}(\hat{J}^{\dagger}\hat{J})_{lm}\text{.}%
\]
In a more compact notation, it is customary to take the columns of the density
matrix, and pile them one below the previous one, transforming the matrix into
a vector%
\begin{equation}
\vec{\rho}=\operatorname{col}(\rho_{11},\rho_{21},...,\rho_{d1},\rho_{12}%
,\rho_{22},...,\rho_{d2},...,\rho_{1d},\rho_{2d},...,\rho_{dd}),
\end{equation}
and the multidimensional array $\{L_{nm;kl}\}_{nmkl=1,2,...,d}$ into a matrix
$\mathbb{L}$, so that the previous equation is turned into the linear system%
\begin{equation}
\frac{d\vec{\rho}}{dt}=\mathbb{L}\vec{\rho}, \label{VectorizedMasterEq}%
\end{equation}
with solution%
\begin{equation}
\vec{\rho}(t)=e^{\mathbb{L}t}\vec{\rho}(0)\text{.}%
\end{equation}
It is interesting to note the correspondence between the elements of the
density matrix, and the elements of its \textit{vectorized} form: $(\vec{\rho
})_{n+(m-1)d}=\rho_{nm}$. In the next section we will see that these
expressions are much more than just a convenient way of reordering things.

From a practical point of view, if the Hilbert space dimension is not too
large, then $e^{\mathbb{L}t}$ can be efficiently evaluated, and the main
problem consists in how to write the matrix $\mathbb{L}$ in an easy way,
starting from the expression of the Liouvillian in the master equation
(\ref{MasterEq}). This is where superspace enters into play, and we will
explain in the next section how it allows for a simple way of coding
$\mathbb{L}$ in the computer (Matlab in particular).

Let us now pass to discuss the important concept of \textit{steady state}. For
problems without selective measurements involved in the system+environment,
the master equation must map states into states; this means that it is a trace
preserving differential map, so that the condition $\mathrm{tr}\{\hat{\rho
}\}=\rho_{11}+\rho_{22}+...+\rho_{dd}=1$ is satisfied at all times, and hence
the equations of (\ref{VectorizedMasterEq}) are not independent, but satisfy
the constrain $\dot{\rho}_{11}+\dot{\rho}_{22}+...+\dot{\rho}_{dd}=0$, the dot
denoting time-derivative. This makes the rows (or columns) of the matrix
$\mathbb{L}$ linearly dependent, what ensures $\det\{\mathbb{L}\}=0$ and hence
that it exists at least one eigenvector with zero eigenvalue, which we will
denote by $\vec{\rho}_{0}$, satisfying $\mathbb{L}\vec{\rho}_{0}=\vec{0}$,
where $\vec{0}$ is a vector of zeros. The corresponding operator $\hat{\rho
}_{0}$ is called the \textit{steady state} of the system, since in the absence
of any other zero eigenvalue, this is the state towards which the system tends
to as time evolves (the trace-preservation condition ensures also that all the
other eigenvalues have negative real part, and hence for long times only
$\vec{\rho}_{0}$ survives).

One useful way of finding this steady state is as follows. First, given its
defining equation $\mathbb{L}\vec{\rho}_{0}=\vec{0}$, where $\vec{0}$ is a
vector of zeros, we replace one of the equations coming from the evolution
equation of some some chosen diagonal element $\rho_{ll}$ by the normalization
condition $\gamma(\rho_{11}+\rho_{22}+...+\rho_{dd})=\gamma$, which can be
additionally multiplied by any number $\gamma$, what is sometimes useful for
numerical purposes; this means replacing the row number $l+(l-1)d$ of
$\mathbb{L}$ by a vector of $\gamma$'s in the elements multiplying the
diagonal elements of $\hat{\rho}_{0}$, obtaining a new matrix $\mathbb{L}_{0}%
$, also replacing the $\vec{0}$ vector by a vector $\vec{w}_{0}$ containing a
single non-zero entry $\gamma$ at position $l+(l-1)d$. The steady state can
then be found by solving the linear system $\mathbb{L}_{0}\vec{\rho}_{0}%
=\vec{w}_{0}$, for example by inversion: $\vec{\rho}_{0}=\mathbb{L}_{0}%
^{-1}\vec{w}_{0}$. Later we will learn how to do this explicitly in Matlab.

\section{The master equation in superspace}

The space where the density matrix is turned into a vector and the Liouvillian
into a matrix is usually called \textit{superspace}. Having operators as its
elements, superspace can be defined formally as the tensor product of the
Hilbert space and its dual, which indeed has vector space structure when
endorsed with the trace product. As we will see in the next section, this
gives us a simple way of representing superoperators by using simple tools of
computer programs such as the Kronecker product, which is a built-in operation
in both in Matlab and Mathematica. Instead of recalling the dual space, it is
computationally more convenient to define superspace in a slightly simpler
way: given an operator $\hat{O}$ with matrix elements $\{O_{nm}%
\}_{n,m=1,2,...,d}$, we just associate to every index a fictitious
$d$-dimensional Hilbert space $\mathcal{F}$ with basis $\{|n\rangle
\}_{n=1,2,...,d}$, in which we describe the operator as a vector%
\begin{equation}
|\hat{O}\rangle=\sum_{nm=1}^{d}O_{nm}|n\rangle\otimes|m\rangle.
\end{equation}
We will say that $|\hat{O}\rangle$ is the abstract superspace representation
of the operator $\hat{O}$. Note that its representation in the basis
$\{|n\rangle\otimes|m\rangle\}_{n,m=1,2,...,d}$ of superspace is the vector
$\vec{O}$ obtained by piling up the columns of the matrix formed by the
elements $O_{nm}$ one below the previous one, but only provided that the order
of the superspace basis $\{\widetilde{|p\rangle}\}_{p=1,2,...,d^{2}}$ is
chosen as\footnote{Convince yourself of this fact through some simple
examples. For example, the first element of the second column, $O_{12}$,
should correspond to the $d+1$ element in the superspace vector, $(\vec
{O})_{d+1}$, and this is precisely what the map $(n,m)\rightarrow n+(m-1)d$
provides.} $\{\widetilde{|n+(m-1)d\rangle}=|n\rangle\otimes|m\rangle
\}_{n,m=1,2,...,d}$.

Consider now two operators $\hat{A}$ and $\hat{B}$, and a superoperator
$\mathcal{S}$ acting on an operator $\hat{O}$ as $\mathcal{S}[\hat{O}]=\hat
{A}\hat{O}\hat{B}=\sum_{nmkl=1}^{d}O_{nm}A_{kn}B_{ml}|k\rangle\langle l|$,
expression which reads in superspace as
\begin{equation}
|\mathcal{S}[\hat{O}]\rangle=\sum_{nmkl=1}^{d}O_{nm}A_{kn}B_{ml}%
|k\rangle\otimes|l\rangle.
\end{equation}
Taking into account that $\hat{A}$ and $\hat{B}$ are operators defined in the
original $d$-dimensional Hilbert space $\mathcal{H}$, so that they act on
basis elements of the new fictitious spaces $\mathcal{F}$ in the usual way
(e.g., $\hat{A}|n\rangle=\sum_{k=1}^{d}A_{kn}|k\rangle$), we can alternatively
write
\begin{equation}
|\mathcal{S}[\hat{O}]\rangle=\sum_{nm=1}^{d}O_{nm}(\hat{A}|n\rangle\otimes
\hat{B}^{T}|m\rangle)=(\hat{A}\otimes\hat{B}^{T})|\hat{O}\rangle,
\end{equation}
showing that in superspace the action of operators on the left (right),
corresponds to actions of the (transpose) operator on the first (second)
fictitious Hilbert space.

Hence, in superspace the master equation can be written as%
\begin{equation}
\frac{d}{dt}|\hat{\rho}\rangle=[-\mathrm{i}(\hat{H}\otimes\hat{I}%
)+\mathrm{i}(\hat{I}\otimes\hat{H}^{T})+2\Gamma(\hat{J}\otimes\hat{J}^{\ast
})-\Gamma(\hat{J}^{\dagger}\hat{J}\otimes\hat{I})-\Gamma(\hat{I}\otimes\hat
{J}^{T}\hat{J}^{\ast})]|\hat{\rho}\rangle,
\end{equation}
where $\hat{I}$ is the identity operator.

\section{Implementing the superspace ideas in Matlab\label{MatlabImp}}

Let's pass now to discuss how to implement the previous ideas in one
particular program, Matlab, although similar tricks can be used in
Mathematica, for example.

First, let us remind the notation, since I wouldn't be surprised if everyone
is lost on it at this point; to complicate things a bit, we will even need to
introduce some more. The (column) vector representation of the basis elements
$\{|n\rangle\}_{n=1,2,...,d}$ of the Hilbert spaces $\mathcal{H}$ or
$\mathcal{F}$ will be denoted by $\{\mathbf{v}_{n}\}_{n=1,2,...,d}$, with
self-representation elements $\left(  \mathbf{v}_{n}\right)  _{m}=\langle
m|n\rangle=\allowbreak\delta_{mn}$. Similarly, the superspace basis
$\{\widetilde{|p\rangle}\}_{p=1,2,...,d^{2}}$ will have a vector
representation $\{\vec{v}_{p}\}_{p=1,2,...,d^{2}}$, with self representation
$\left(  \vec{v}_{p}\right)  _{q}=\delta_{qp}$. Given an operator $\hat{O}$,
its matrix elements are denoted by $O_{nm}=$ $\langle n|\hat{O}|m\rangle$, and
the $d\times d$ matrix that they form by $\mathbf{O}$, which is nothing but
the matrix representation of the operator in the chosen basis. The
`vectorized' form of this matrix, that is, the vector formed by piling up the
columns of the matrix one below the next, is denoted by $\vec{O}$, and, as
explained above, it can be seen as the representation of the operator in
superspace, which we will still denote as $|\hat{O}\rangle$, provided that the
basis elements of superspace are ordered in the proper way. Hence, summing up,
$\hat{O}$ is the abstract notation for the operator acting on the original
space $\mathcal{H}$ and $|\hat{O}\rangle$ the one for the operator defined in
superspace $\mathcal{F}\otimes\mathcal{F}$, with corresponding matrix and
vector representations $\mathbf{O}$ and $\vec{O}$, respectively. As for
superoperators, take the Liouvillian as an example, we will refer to them as
$\mathcal{L}$ in calligraphic font\footnote{Not to confuse with the notation
for Hilbert spaces, for which we also use calligraphic font, but it should be
clear from the context when we mean one or the other.} when talking about them
in an abstract way, and $\mathbb{L}$ in blackboard font when referring to
their matrix representation in superspace. For example, the right hand side of
master equation (\ref{MasterEq}) reads $\mathcal{L}[\hat{\rho}]$ in an
abstract way, and as $\mathbb{L}\vec{\rho}$ once represented in superspace.

Let's start from the basics of coding things in Matlab. The matrix
representation of an operator $\hat{O}$ can be written in Matlab (known their
matrix elements $O_{nm}$) as
\begin{equation}
\mathbf{O}=[O_{11},O_{12},...,O_{1d};O_{21},O_{22},...,O_{2d};...;O_{d1}%
,O_{d2},...,O_{dd}],
\end{equation}
where the commas can be replaced by a space. If we want to work with sparse
matrices to save memory (useful for Hilbert spaces with large dimension, e.g.,
$d>10$), we can do so by replacing the matrix $\mathbf{O}$ by\footnote{From
now on, Matlab functions will be highlighted by using typewriter font.}
\texttt{sparse}($\mathbf{O}$); once in sparse form, we can always come back to
the non-sparse one as \texttt{full}($\mathbf{O}$).

Let's talk about basic matrix operations. Element $O_{nm}$ is accessed as
$\mathbf{O}(n,m)$, while the whole column $m$ can be accessed as
$\mathbf{O}(:,m)$, and similarly for row $n$, $\mathbf{O}(n,:)$. We can access
its $n$-th diagonal as $\mathtt{diag}(\mathbf{O},n)$ which generates a column
vector with the desired diagonal. Given another matrix $\mathbf{Q}$,
$\mathbf{O\pm Q}$ is the matrix sum or difference, $\mathbf{O}$*$\mathbf{Q}$
is the matrix multiplication, $\mathbf{O}.$*$\mathbf{Q}$ is the
element-by-element multiplication, $\mathbf{O}\backslash\mathbf{Q}$ is the
matrix multiplication of $\mathbf{O}^{-1}$ and $\mathbf{Q}$, and
$\mathbf{O}/\mathbf{Q}$ is the matrix multiplication of $\mathbf{O}$ and
$\mathbf{Q}^{-1}$, where the inverse of $\mathbf{O}$ can also be obtained as
\texttt{inv}$(\mathbf{O})$ (but it is not recommended by Matlab, since it is
slower than $1/\mathbf{O}$ or $\mathbf{O}\backslash1$). We can find the
determinant and trace as \texttt{det}$(\mathbf{O})$ and \texttt{trace}%
$(\mathbf{O})$. The Hermitian conjugate of $\mathbf{O}$ is obtained in Matlab
as $\mathbf{O}^{\prime}$, while $\mathbf{O}.^{\prime}$ generates the transpose
of $\mathbf{O}$. The exponential matrix is obtained as \texttt{expm}%
$(\mathbf{O})$, while \texttt{exp}$(\mathbf{O})$ just exponentiates the
elements of the matrix individually. As for the eigensystem, $\mathtt{eig}%
(\mathbf{O})$ generates a vector with the eigenvalues of $\mathbf{O}$, while
if we write $[\mathbf{V},\mathbf{D}]=$\texttt{eig}$(\mathbf{O})$, the
eigenvectors are codified as columns of $\mathbf{V}$ and $\mathbf{D}$ is a
diagonal matrix containing the corresponding eigenvalues. This operation
cannot be used when $\mathbf{O}$ is sparse, in which case we need to use
\texttt{eigs}$(\mathbf{O})$, which by default gives the 6 eigenvalues with
largest magnitude; if we want a different number of eigenvalues, say $N$, we
can write \texttt{eigs}$(\mathbf{O},N,$`$xy$'$)$, where $x=$ \texttt{L}
(\texttt{S}) means that we want the eigenvalues with the largest (smallest)
magnitude if $y=$ \texttt{M}, or real part if $y=$ \texttt{R}. It is very
useful to type \texttt{help F} in Matlab's command window to get more info
about some function \texttt{F} (for example, try \texttt{help eig} to find
what else can be done with \texttt{eig}).

The vectorized form of the operator is obtained as $\vec{O}=\mathbf{O}(:)$.
This is one of the reasons why we chose to pile columns instead of rows when
vectorizing matrix representations of operators: in Matlab this operation is
written with a single order, while in the case of rows, we would need to first
transpose the matrix, and then give the order. On the other hand, given the
matrix in vectorized form, we can always bring it back to matrix form as
$\mathbf{O}=$ \texttt{reshape}$(\vec{O},d,d)$.

The matrix representation of the identity operator $\hat{I}$ can be written as
$\mathbf{I}=$ \texttt{eye}$(d)$, or $\mathbf{I}=$ \texttt{speye}$(d)$ when
working with sparse matrices. The basis vector $\mathbf{v}_{n}$ can be defined
as $\mathbf{v}_{n}=\mathbf{I}(:,n)$. Similarly, defining the identity matrix
in $d^{2}$ dimensions $\mathbf{\tilde{I}}=$ \texttt{eye}$(d^{2})$, the basis
vector $\vec{v}_{p}$ in superspace is obtained as $\vec{v}_{p}=\mathbf{\tilde
{I}}(:,p)$.

Let's move on to the tensor product operation. It is customary in quantum
mechanics to represent the tensor product of two operators or vectors as the
Kronecker product of their representations. In particular, given the matrix
representations $\mathbf{A}$ and $\mathbf{B}$ of two operators $\hat{A}$ and
$\hat{B}$, their Kronecker product is defined as%
\begin{equation}
\left(
\begin{array}
[c]{cccc}%
A_{11}\mathbf{B} & A_{12}\mathbf{B} & \cdots & A_{1d}\mathbf{B}\\
A_{21}\mathbf{B} & A_{22}\mathbf{B} & \cdots & A_{2d}\mathbf{B}\\
\vdots & \vdots & \ddots & \vdots\\
A_{d1}\mathbf{B} & A_{d2}\mathbf{B} & \cdots & A_{dd}\mathbf{B}%
\end{array}
\right)  ,
\end{equation}
and this is usually the representation chosen for the tensor product operator
$\hat{A}\otimes\hat{B}$. However, it is important to understand that this is
just one possible representation of the tensor product, corresponding to one
particular ordering of the tensor product basis $\{|n\rangle\otimes
|m\rangle\}_{n,m=1,2,...,d}$ in the composite Hilbert space (superspace in our
case). More concretly, note that according to the previous definition, given
the vector representation $\mathbf{v}_{n}$\ of the basis element $|n\rangle
\in\mathcal{H}$ (or $\mathcal{F}$), the Kronecker product representation of
the basis element $|n\rangle\otimes|m\rangle\in\mathcal{H\otimes H}$ (or
$\mathcal{F\otimes F}$) generates a vector with a single nonzero entry at
position $m+(n-1)d$, that is, the superspace basis vector\footnote{Convince
yourself of this fact by working out some examples, e.g., in dimension 3
($d=3$), the Kronecker product of $\mathbf{v}_{1}=\operatorname{col}(1,0,0)$
and $\mathbf{v}_{2}=\operatorname{col}(0,1,0)$, generates the vector $\vec
{v}_{2}=\operatorname{col}(0,1,0,0,0,0,0,0,0)$, while the Kronecker product of
$\mathbf{v}_{2}$ and $\mathbf{v}_{1}$ generates $\vec{v}_{5}%
=\operatorname{col}(0,0,0,0,1,0,0,0,0)$. These examples coincide precisely
with the mapping $(n,m)\rightarrow m+(n-1)d$.} $\vec{v}_{m+(n-1)d}$. However,
as explained in the previous section, we would like to associate instead the
superspace basis vector $\vec{v}_{n+(m-1)d}$ to the tensor product basis
element $|n\rangle\otimes|m\rangle$, what means that we will not be using the
usual Kronecker product representation of the tensor product, but one in
reversed order: given the matrix representations $\mathbf{A}$ and $\mathbf{B}$
of two operators $\hat{A}$ and $\hat{B}$, the representation of their tensor
product $\hat{A}\otimes\hat{B}$ is taken as their Kronecker product in
reversed order, that is,%
\begin{equation}
\left(
\begin{array}
[c]{cccc}%
B_{11}\mathbf{A} & B_{12}\mathbf{A} & \cdots & B_{1d}\mathbf{A}\\
B_{21}\mathbf{A} & B_{22}\mathbf{A} & \cdots & B_{2d}\mathbf{A}\\
\vdots & \vdots & \ddots & \vdots\\
B_{d1}\mathbf{A} & B_{d2}\mathbf{A} & \cdots & B_{dd}\mathbf{A}%
\end{array}
\right)  .
\end{equation}
With this choice, the representation of the basis element $|n\rangle
\otimes|m\rangle$ corresponds to $\vec{v}_{n+(m-1)d}$ as we wanted to.

The Kronecker product is already implemented in Matlab through the operation
\texttt{kron} (and same in Mathematica), which preserves the sparse character
of the matrices. Hence, we can generate the vector representation of
$\widetilde{|n+(m-1)d\rangle}=|n\rangle\otimes|m\rangle$, by applying the
\texttt{kron} operation in the reversed order $\vec{v}_{n+(m-1)d}=$
\texttt{kron}$(\mathbf{v}_{m},\mathbf{v}_{n})$.

Consider now two operators $\hat{A}$ and $\hat{B}$, and a superoperator
$\mathcal{S}$ which acts on a third operator $\hat{O}$ as $\mathcal{S}[\hat
{O}]=\hat{A}\hat{O}\hat{B}$. As explained in the previous section, in
superspace this is rewritten as $|\mathcal{S}[\hat{O}]\rangle=(\hat{A}%
\otimes\hat{B}^{T})|\hat{O}\rangle$ in an abstract way, expression which can
be represented in the basis of superspace $\{\widetilde{|p\rangle
}\}_{p=1,2,...,d^{2}}$ as $\mathbb{S}\vec{O}=$ \texttt{kron}$(\mathbf{B}%
.^{\prime},\mathbf{A})$*$\mathbf{O}(:)$ in Matlab code. Hence, the matrix
representation of $\mathcal{S}$ in superspace is written in Matlab as
$\mathbb{S}=$ \texttt{kron}$(\mathbf{B}.^{\prime},\mathbf{A})$.

Let me remark that the choice of using the reversed \texttt{kron} order for
the representation of the tensor product has been made for convenience in
Matlab (to create the vectorized form of any operator with a single
instruction, and for more things that will appear in the next section when
dealing with composite Hilbert spaces). However, in other languages such as
Mathematica, it can be better to stick to the traditional Kronecker product
representation of the tensor product. But above all, what is important to
understand what one is doing, and hence I strongly encourage the reader to
think deeply about this, and play with some examples to interiorize this
tricky point.

Hence, as promised, the matrix representation of the Liouvillian superoperator
admits a very simple coding in Matlab:%
\begin{equation}
\mathbb{L}=-\text{\texttt{1i}*\texttt{kron}}(\mathbf{I},\mathbf{H}%
)+\text{\texttt{1i}*\texttt{kron}}(\mathbf{H}.^{\prime},\mathbf{I}%
)+2\text{*}\Gamma\text{*\texttt{kron}}(\text{\texttt{conj}}(\mathbf{J)}%
,\mathbf{J})-\Gamma\text{*\texttt{kron}}(\mathbf{I},\mathbf{J}^{\prime
}\text{*}\mathbf{J})-\Gamma\text{*\texttt{kron}}(\mathbf{J}.^{\prime
}\text{*\texttt{conj}}(\mathbf{J}),\mathbf{I}), \label{LinMatlab}%
\end{equation}
where \texttt{1i} is the proper way of writing the imaginary unit in Matlab,
while \texttt{conj}$($\texttt{z}$)$ is how the complex conjugate of \texttt{z}
looks in Matlab.

Once we have the Liouvillian superoperator, the next issue concerns finding
the steady state of the system. If the Hilbert space dimension is not too
large, we can try diagonalizing $\mathbb{L}$ fully using $[\mathbb{V}%
,\mathbb{D}]=$ \texttt{eig}$(\mathbb{L})$. To access the steady state we can
just find the index of the zero eigenvalue and get the corresponding column of
$\mathbb{V}$, or proceed in a more elegant and automatic way, by sorting the
order in which the eigenvalues appear. In particular, given the vector of
eigenvalues $\boldsymbol{\lambda}=$ \texttt{diag}$(\mathbb{D})$, we generate a
vector $\mathbf{y}$ containing the indices of the permutation which we need to
apply to reorder the eigenvectors in decreasing real part as $[\mathbf{x}%
,\mathbf{y}]=$ \texttt{sort}$($\texttt{real}$(\boldsymbol{\lambda}%
),$`\texttt{descend}'$)$, where we additionally get the vector of sorted real
parts $\mathbf{x}$, which we won't use; once we have $\mathbf{y}$, we can sort
the eigensystem as $\mathbb{V}=\mathbb{V}(:,\mathbf{y})$ and
$\boldsymbol{\lambda}=\boldsymbol{\lambda}(\mathbf{y})$, and the first column
of the sorted $\mathbb{V}$ should correspond now to the steady state
$\vec{\rho}_{0}=\mathbb{V}(:,1)$, most likely requiring the additional
normalization $\vec{\rho}_{0}=\vec{\rho}_{0}/$\texttt{trace}$($%
\texttt{reshape}$(\vec{\rho}_{0},d,d))$ to ensure it has unit trace. Now, for
larger size problems, we will need to use sparse matrices, in which case the
simplest way of finding the steady state would be as $\vec{\rho}_{0}=$
\texttt{eigs}$(\mathbb{L},1,$`\texttt{LR}'$)$, which might require additional
normalization as in the previous line. The density matrix of the steady state
can then be found as $\boldsymbol{\rho}_{0}=$ \texttt{reshape}$(\vec{\rho}%
_{0},d,d)$.

Even though in most cases the previous way of finding $\vec{\rho}_{0}$ is
enough, it is interesting to know how to implement the method which we
introduced at the end of the first section, which consisted in replacing one
equation of $\mathbb{L}\vec{\rho}_{0}=\vec{0}$ by the normalization condition
$\gamma(\rho_{11}+\rho_{22}+...+\rho_{dd})=\gamma$, where $\gamma>0$ is a
parameter which we can choose as we wish. This is easily done in Matlab as
follows (there are indeed many different ways of doing this, here I just pick
the one I find simplest and most direct to code). First, we pick the index $l$
of the diagonal element $\rho_{ll}$ whose equation we want to replace, with
corresponding superindex $s_{l}=l+(l-1)d$. Then, we just define the matrix
$\mathbb{L}_{0}=\mathbb{L}$, and replace the corresponding row as
$\mathbb{L}_{0}(s_{l},:)=\gamma$*$\mathbf{I}(:)$, which simply puts $\gamma$
on the entries multiplying the diagonal elements of the density matrix. Then
we define the vector $\vec{w}_{0}$ with a single $\gamma$ on element $s_{l}$,
which indeed corresponds to the superspace basis element $\vec{v}_{s_{l}}$
multiplied by $\gamma$, that is, we can simply code it as $\vec{w}_{0}=\gamma
$*$\mathbf{\tilde{I}}(:,s_{l})$. Once we have done this, $\vec{\rho}_{0}$ can
be found as $\vec{\rho}_{0}=$ \texttt{inv}$(\mathbb{L}_{0})$*$\vec{w}_{0}$, or
by asking Matlab to solve the linear system $\mathbb{L}_{0}\vec{\rho}_{0}%
=\vec{w}_{0}$ as $\vec{\rho}_{0}=$ \texttt{linsolve}$(\mathbb{L}_{0},\vec
{w}_{0})$, which uses LU factorization. Note that we need to check that
\texttt{det}$(\mathbb{L}_{0})\neq0$, as otherwise we will have degenerate
steady states, and the method will fail.

\section{Dealing with composite Hilbert spaces\label{Composite}}

Everything we introduced up to now is general, in the sense that it applies to
a system with any Hilbert space $\mathcal{H}$. Here I want to discuss some
special features that appear when the system is a composition of $N$ simpler
subsystems (two- or three-level systems, harmonic oscillators, etc...), in
which case the Hilbert space has the tensor product structure $\mathcal{H}%
=\mathcal{H}_{1}\otimes\mathcal{H}_{2}\otimes...\otimes\mathcal{H}_{N}$. This
is the scenario that we usually find in quantum optics, where typical systems
are composed of atoms (maybe artificial such as superconducting qubits or
quantum dots), and/or photonic, phononic, or motional modes.

Given a basis $\{|n_{j}\rangle\}_{n_{j}=1,2,...,d_{j}}$ of subspace
$\mathcal{H}_{j}$ with dimension $d_{j}$, a basis of the full Hilbert space
can be built as $\{|n_{1}\rangle\otimes|n_{2}\rangle\otimes...\otimes
|n_{N}\rangle\}_{n_{j}=1,2,...,d_{j}}\equiv\{|n\rangle\}_{n=1,2,...,d}$, where
$d=d_{1}\times d_{2}\times...\times d_{N}$ is the dimension of the complete
Hilbert space. Hence, each value of the index $n$ which we were using in the
previous sections corresponds now to some multi-index $n_{1}n_{2}...n_{N}$
labeling the basis elements of the Hilbert space of each subsystem. The point
is that in many cases (for example when wanting to evaluate partial traces or
transpositions) it is important to keep track of all the indices, and here I
want to discuss how to deal with these issues by using efficient computational tools.

We have already encountered tensor products before when building the
superspace, and we saw how to code them efficiently using the Kronecker
product; we will again use the \texttt{kron} operation to deal with composite
Hilbert spaces, but with a few subtle points. Let us start with the simplest
structure in the composite Hilbert space: let's represent its basis. Consider
again the basis $\{|n_{j}\rangle\}_{n_{j}=1,2,...,d_{j}}$ of subspace
$\mathcal{H}_{j}$, and define the identity matrix of the corresponding
dimension, $\mathbf{I}^{(j)}=$ \texttt{eye}$(d_{j})$, from which we build the
vector representation of $|n_{j}\rangle$ as $\mathbf{v}_{n_{j}}^{(j)}%
=\mathbf{I}^{(j)}(:,n_{j})$. Then, we will represent a basis element $n$ of
the complete Hilbert space corresponding to some multi-index $n_{1}%
n_{2}...n_{N}$, that is, $|n\rangle=|n_{1}\rangle\otimes|n_{2}\rangle
\otimes...\otimes|n_{N}\rangle$, as the vector $\mathbf{v}_{n}=$
\texttt{kron}$(\mathbf{v}_{n_{N}}^{(N)},...,$\texttt{kron}$(\mathbf{v}_{n_{3}%
}^{(3)},$\texttt{kron}$(\mathbf{v}_{n_{2}}^{(2)},\mathbf{v}_{n_{1}}%
^{(1)}))...)$, where we use again a reversed order in the \texttt{kron}
operations for future convenience, see the next paragraph. Taking into account
that we still want to define $\mathbf{v}_{n}$ as a vector with $d-1$ zeros and
a one at position $n$, which is the natural self-representation of the basis
elements in the complete Hilbert space, the previous definition fixes the
relation between $n$ and the multi-index $n_{1}n_{2}...n_{N}$ to%
\begin{equation}
n=n_{1}+(n_{2}-1)d_{1}+(n_{3}-1)d_{1}d_{2}+...+(n_{N}-1)d_{1}d_{2}%
...d_{N-1}\text{.} \label{IndxToSuperIndx}%
\end{equation}

Consider now a pure state $|a\rangle=\sum_{n=1}^{d}a_{n}|n\rangle$ in the
complete Hilbert space. As explained, it would be useful to be able to move
between this expression, and the one making explicit reference to the indices
of each subspace, that is, $|a\rangle=\sum_{n_{1}=1}^{d_{1}}\sum_{n_{2}%
=1}^{d_{2}}...\sum_{n_{N}=1}^{d_{N}}a_{n_{1}n_{2}...n_{N}}|n_{1}\rangle
\otimes|n_{2}\rangle\otimes...\otimes|n_{N}\rangle$. This is very easy to do
in Matlab once we have made all the previous definitions. In particular, given
the (column) vector representation of the state $\mathbf{a}=\operatorname{col}%
(a_{1},a_{2},...,a_{d})$ or better $\mathbf{a}=[a_{1};a_{2};...;a_{d}]$ in
Matlab code, we can transform it into a multidimensional array as
$\mathbf{\ddot{a}}=$ \texttt{reshape}$(\mathbf{a},d_{1},d_{2},...,d_{N})$,
from which $a_{n_{1}n_{2}...n_{N}}$ is simply accessed as $a_{n_{1}%
n_{2}...n_{N}}=\mathbf{\ddot{a}}(n_{1},n_{2},...,n_{N})$; in the following we
will use the double dot on top of the bold-faced symbol to denote that it is a
multidimensional array. The multidimensional array can be transformed back to
its vector form in the full Hilbert space just vectorizing it as
$\mathbf{a}=\mathbf{\ddot{a}}(:)$. Note that the dimensions $(d_{1}%
,d_{2},...,d_{N})$ used to reshape the vector and the indices $(n_{1}%
,n_{2},...,n_{N})$ of the multidimensional array, follow the intuitive order
that one would assign; this is thanks to using the reversed order in the
\texttt{kron} operation, and would not be the case if we would have chosen the
intuitive one.

At this point it should be clear that, given substates $\{|a^{(j)}\rangle
=\sum_{n_{j}=1}^{d_{j}}a_{n_{j}}^{(j)}|n_{j}\rangle\in\mathcal{H}%
_{j}\}_{j=1,2,...,N}$ with vector representation $\mathbf{a}^{(j)}%
=\operatorname{col}(a_{1}^{(j)},a_{2}^{(j)},...,a_{d_{j}}^{(j)})$ or
$\mathbf{a}^{(j)}=[a_{1}^{(j)};a_{2}^{(j)};...;a_{d_{j}}^{(j)}]$ in Matlab
code, the vector representation of the tensor product state $|a\rangle
=|a^{(1)}\rangle\otimes|a^{(2)}\rangle\otimes...\otimes|a^{(N)}\rangle$ can be
obtained in Matlab as $\mathbf{a}=$ \texttt{kron}$(\mathbf{a}^{(N)}%
,...,$\texttt{kron}$(\mathbf{a}^{(3)},$\texttt{kron}$(\mathbf{a}%
^{(2)},\mathbf{a}^{(1)}))...)$, with elements $a_{n}=a_{n_{1}}^{(1)}a_{n_{2}%
}^{(2)}...a_{n_{N}}^{(N)}$, $n$ given by (\ref{IndxToSuperIndx}).

We see then that dealing with vectors is not so difficult. Now let's consider
operators, which are a bit more tricky. Let's start with a simple
generalization of what we did for vectors. Consider an operator $\hat{O}%
=\sum_{nm=1}^{d}O_{nm}|n\rangle\langle m|$ in the complete Hilbert space.
Given its matrix representation $\mathbf{O}$ in Matlab (with elements $O_{nm}%
$) we would like to be able to retrieve the multi-index elements
$O_{n_{1}n_{2}...n_{N};m_{1}m_{2}...m_{N}}$ defined from%
\begin{equation}
\hat{O}=\sum_{n_{1},m_{1}=1}^{d_{1}}\sum_{n_{2},m_{2}=1}^{d_{2}}...\sum
_{n_{N},m_{N}=1}^{d_{N}}O_{n_{1}n_{2}...n_{N};m_{1}m_{2}...m_{N}}%
(|n_{1}\rangle\otimes|n_{2}\rangle\otimes...\otimes|n_{N}\rangle)(\langle
m_{1}|\otimes\langle m_{2}|\otimes...\otimes\langle m_{N}|).
\end{equation}
Similarly to vectors, this can be done by redefining $\mathbf{O}$ as a
multi-dimensional array $\mathbf{\ddot{O}}=$ \texttt{reshape}$(\mathbf{O}%
,d_{1},d_{2},...,d_{N},d_{1},d_{2},...,d_{N})$, from which we can then get the
desired multi-index elements as $O_{n_{1}n_{2}...n_{N};m_{1}m_{2}...m_{N}%
}=\mathbf{\ddot{O}}(n_{1},n_{2},...,n_{N},m_{1},m_{2},...,m_{N})$. We can
always go back to the original matrix representation in the complete Hilbert
space by reshaping the multidimensional array as $\mathbf{O}=$
\texttt{reshape}$(\mathbf{\ddot{O}},d,d)$.

Sometimes it is useful to have access to a different set of multi-index
elements $O_{n_{1}m_{1};n_{2}m_{2};...;n_{N}m_{N}}$ defined by%
\begin{equation}
\hat{O}=\sum_{n_{1},m_{1}=1}^{d_{1}}\sum_{n_{2},m_{2}=1}^{d_{2}}...\sum
_{n_{N},m_{N}=1}^{d_{N}}O_{n_{1}m_{1};n_{2}m_{2};...;n_{N}m_{N}}|n_{1}%
\rangle\langle m_{1}|\otimes|n_{2}\rangle\langle m_{2}|\otimes...\otimes
|n_{N}\rangle\langle m_{N}|.
\end{equation}
For this, the best is first building the multidimensional array $\mathbf{\ddot
{O}}$ as we explained above, and then use the extremely useful
\texttt{permute} operation, which allows to permute indices of
multidimensional arrays. In particular, defining another multidimensional
array $\mathbf{\ddddot{O}}=$ \texttt{permute}$(\mathbf{\ddot{O}}%
,[1,N+1,2,N+2,...,N,2N])$, we then access the desired multi-index elements as
$O_{n_{1}m_{1};n_{2}m_{2};...;n_{N}m_{N}}=\mathbf{\ddddot{O}}(n_{1}%
,m_{1},n_{2},m_{2},...,n_{N},m_{N})$. In the following we will use the
notation $\mathbf{\ddot{O}}$ for the multidimensional array corresponding to
the order $O_{n_{1}n_{2}...n_{N};m_{1}m_{2}...m_{N}}$ of the multi-index
elements, and the notation $\mathbf{\ddddot{O}}$ for the one corresponding to
the $O_{n_{1}m_{1};n_{2}m_{2};...;n_{N}m_{N}}$ order.

Imagine now that we are given a set of operators $\{\hat{O}_{j}%
\}_{j=1,2,...,N}$ acting on the subspaces $\{\mathcal{H}_{j}\}_{j=1,2,...,N}$,
with corresponding matrix representations $\{\mathbf{O}_{j}\}_{j=1,2,...,N}$.
Our goal now is finding the different representations of the operator $\hat
{O}=\hat{O}_{1}\otimes\hat{O}_{2}\otimes...\otimes\hat{O}_{N}$ acting on the
complete Hilbert space. Our starting point will be the Kronecker product
$\mathbf{\tilde{O}}=$ \texttt{kron}$(\mathbf{O}_{N},...,$\texttt{kron}%
$(\mathbf{O}_{3},$\texttt{kron}$(\mathbf{O}_{2},\mathbf{O}_{1}))...)$; even
though this is a $d\times d$ matrix containing the elements of the
representation of $\hat{O}$ in the complete basis $\{|n\rangle\}_{n=1,2,...,d}%
$, it is easy to see that these elements are not ordered in the right way, and
hence, in order for $\mathbf{\tilde{O}}$ to coincide with the proper matrix
representation of $\hat{O}$ we need to reorder its elements. To this aim, we
first build the multidimensional array $\mathbf{\ddddot{O}}=$ \texttt{reshape}%
$(\mathbf{\tilde{O}},d_{1},d_{1},d_{2},d_{2},...,d_{N},d_{N})$ which has
$\mathbf{\ddddot{O}}(n_{1},m_{1},n_{2},m_{2},...,n_{N},m_{N})=O_{n_{1}%
m_{1};n_{2}m_{2};...;n_{N}m_{N}}$ as its elements. Then, we can reorder its
indices as $\mathbf{\ddot{O}}=$ \texttt{permute}$(\mathbf{\ddddot{O}%
},[1,3,...,2N-1,2,4,...,2N])$, creating a multidimensional array which has
$\mathbf{\ddot{O}}(n_{1},n_{2},...,n_{N},m_{1},m_{2},...,m_{N})=O_{n_{1}%
n_{2}...n_{N};m_{1}m_{2}...m_{N}}$ as its elements. Finally, we find the
matrix representation of $\hat{O}$ in the complete Hilbert space as
$\mathbf{O}=$ \texttt{reshape}$(\mathbf{\ddot{O}},d,d)$. Hence, essentially we
have followed the path of the previous paragraphs, but in reverse order. I
hope all the manipulations have served to gain intuition about the operations
\texttt{kron}, \texttt{reshape}, and \texttt{permute}, which allow for
efficient and clean ways of representing Hilbert space objects in the computer.

Finally, I would like to discuss two operations very relevant in the context
of composite Hilbert spaces: the partial transposition and the partial trace
of an operator. Consider an operator $\hat{O}$ acting on the complete Hilbert
space. We define the operator corresponding to a partial transpose of $\hat
{O}$ with respect to subspace $\mathcal{H}_{j}$ as%
\begin{align}
\hat{O}^{T_{j}}  &  =\sum_{n_{1},m_{1}=1}^{d_{1}}...\sum_{n_{j},m_{j}%
=1}^{d_{j}}...\sum_{n_{N},m_{N}=1}^{d_{N}}O_{n_{1}...m_{j}...n_{N}%
;m_{1}...n_{j}...m_{N}}(|n_{1}\rangle\otimes...\otimes|n_{j}\rangle
\otimes...\otimes|n_{N}\rangle)(\langle m_{1}|\otimes...\otimes\langle
m_{j}|\otimes...\otimes\langle m_{N}|)\nonumber\\
&  =\sum_{n_{1},m_{1}=1}^{d_{1}}...\sum_{n_{j},m_{j}=1}^{d_{j}}...\sum
_{n_{N},m_{N}=1}^{d_{N}}O_{n_{1}m_{1};...;m_{j}n_{j};...;n_{N}m_{N}}%
|n_{1}\rangle\langle m_{1}|\otimes...\otimes|n_{j}\rangle\langle m_{j}%
|\otimes...\otimes|n_{N}\rangle\langle m_{N}|.
\end{align}
The different multidimensional array representations of this operator are
easily obtained in Matlab from the ones of the original operator as
$\mathbf{\ddot{O}}^{T_{j}}=$ \texttt{permute}$(\mathbf{\ddot{O}}%
,[1,...,j-1,N+j,...,j,N+j+1,...,2N])$ or $\mathbf{\ddddot{O}}^{T_{j}}=$
\texttt{permute}$(\mathbf{\ddddot{O}},[1,...,2(j-1),2j,2j-1,...,2N])$, and its
matrix representation in the complete Hilbert space is then obtained as
$\mathbf{O}^{T_{j}}=$ \texttt{reshape}$(\mathbf{\ddot{O}}^{T_{j}},d,d)$. This
is intuitively generalized to the case in which we want to perform partial
transposition with respect to several subspaces; for example, if we want to
transpose subspaces $\mathcal{H}_{j}$ and $\mathcal{H}_{l}$ ($j<l$), then
\begin{subequations}
\begin{align}
\mathbf{\ddot{O}}^{T_{j}T_{l}}  &  =\mathtt{permute}(\mathbf{\ddot{O}%
},[1,...,j,N+j,...,l,N+l,...,j,N+j+1...,l,N+l+1,...,2N]),\\
\mathbf{\ddddot{O}}^{T_{j}T_{l}}  &  =\mathtt{permute}(\mathbf{\ddddot{O}%
},[1,...,2(j-1),2j,2j-1,...,2(l-1),2l,2l-1,...,2N]),\\
\mathbf{O}^{T_{j}T_{l}}  &  =\mathtt{reshape}(\mathbf{\ddot{O}}^{T_{j}T_{l}%
},d,d).
\end{align}

As for the partial trace over subspace $\mathcal{H}_{j}$, denoted by
$\mathrm{tr}_{j}\{\hat{O}\}$ or $\allowbreak\hat{O}^{\left\{  j\right\}  }$,
it is defined as the operator%
\end{subequations}
\begin{align}
\hat{O}^{\{j\}}  &  =\sum_{n_{1},m_{1}=1}^{d_{1}}...\sum_{n_{N},m_{N}%
=1}^{d_{N}}\left(  \sum_{n_{j}=1}^{d_{j}}O_{n_{1}m_{1};...;n_{j}%
n_{j};...;n_{N}m_{N}}\right)  |n_{1}\rangle\langle m_{1}|\otimes
...\otimes|n_{j-1}\rangle\langle m_{j-1}|\otimes|n_{j+1}\rangle\langle
m_{j+1}|...\otimes|n_{N}\rangle\langle m_{N}|\\
&  =\sum_{n_{1},m_{1}=1}^{d_{1}}...\sum_{n_{N},m_{N}=1}^{d_{N}}\left(
\sum_{n_{j}=1}^{d_{j}}O_{n_{1}...n_{j}...n_{N};m_{1}...n_{j}...m_{N}}\right)
\nonumber\\
&  \text{ \ \ \ \ \ \ \ \ \ \ \ \ \ \ \ \ \ \ \ \ \ }\times(|n_{1}%
\rangle\otimes...\otimes|n_{j-1}\rangle\otimes|n_{j+1}\rangle\otimes
...\otimes|n_{N}\rangle)(\langle m_{1}|\otimes...\otimes\langle m_{j-1}%
|\otimes\langle m_{j+1}|\otimes...\otimes\langle m_{N}|),\nonumber
\end{align}
that is, as the operators with elements corresponding to the contraction of
the indices associated to the $\mathcal{H}_{j}$ subspace. We can find again
the different representations of this operator efficiently using Matlab's
built-in functions. In particular, we can find the multidimensional array
$\mathbf{\ddot{O}}^{\{j\}}$ corresponding to this operator as follows (we
proceed by sequentially updating its definition): first, we bring the indices
that we want to contract to the end of the array, $\mathbf{\ddot{O}}^{\{j\}}=$
\texttt{permute}$(\mathbf{\ddot{O}}%
,[1,...,j-1,j+1,...,N+j-1,N+j+1,...,2N,j,N+j])$; defining the dimension of the
total Hilbert space after tracing out the desired subspace by $d^{\{j\}}%
=d_{1}\times...\times d_{j-1}\times d_{j+1}\times...\times d_{N}$, we reshape
the previous array as a $d^{\{j\}2}\times d_{j}^{2}$ dimensional matrix,
$\mathbf{\ddot{O}}^{\{j\}}=$ \texttt{reshape}$(\mathbf{\ddot{O}}%
^{\{j\}},d^{\{j\}2},d_{j}^{2})$; given the identity matrix of dimension
$d_{j}$ denoted by $\mathbf{I}^{(j)}=$ \texttt{eye}$(d_{j})$, in terms of the
previous matrix the contraction we are looking for is just $\mathbf{\ddot{O}%
}^{\{j\}}=\mathbf{\ddot{O}}^{\{j\}}$*$\mathbf{I}^{(j)}(:)$; the previous
operation leaves us with a column vector with $d^{\{j\}2}$ components, which
we can finally reshape to give the multidimensional array we are looking for,
$\mathbf{\ddot{O}}^{\{j\}}=$ \texttt{reshape}$(\mathbf{\ddot{O}}^{\{j\}}%
,d_{1},...,d_{j-1},d_{j+1},...,d_{N},d_{1},...,d_{j-1},d_{j+1},...,d_{N})$;
finally, we find the matrix representation of the operator in the (remaining)
complete Hilbert space as $\mathbf{\ddot{O}}^{\{j\}}=$ \texttt{reshape}%
$(\mathbf{\ddot{O}}^{\{j\}},d^{\{j\}},d^{\{j\}})$. Of course, a similar trick
can be done starting from the other multidimensional array $\mathbf{\ddddot
{O}}$; also, the method is straightforwardly generalized to when we want to
trace out several subspaces at once (maybe it's a good thing to try these two
things out as an exercise, to really discover if you understood all these
constructions properly).

Let me finally remark once more that the only reason why we have choosen the
reversed order in the \texttt{kron} operation is to make the coding simpler in
Matlab. In particular, by just sticking to the simple rule \textquotedblleft
every time a tensor product appears, it is coded as the \texttt{kron} product
in the reversed order\textquotedblright, the rest of manipulations are
compactly and intuitively coded in Matlab as shown above, what would not be
the case otherwise.

\section{An example: three-level cascade system interacting with two quantized
optical modes}

\begin{figure}[b]
\includegraphics[width=0.7\columnwidth]{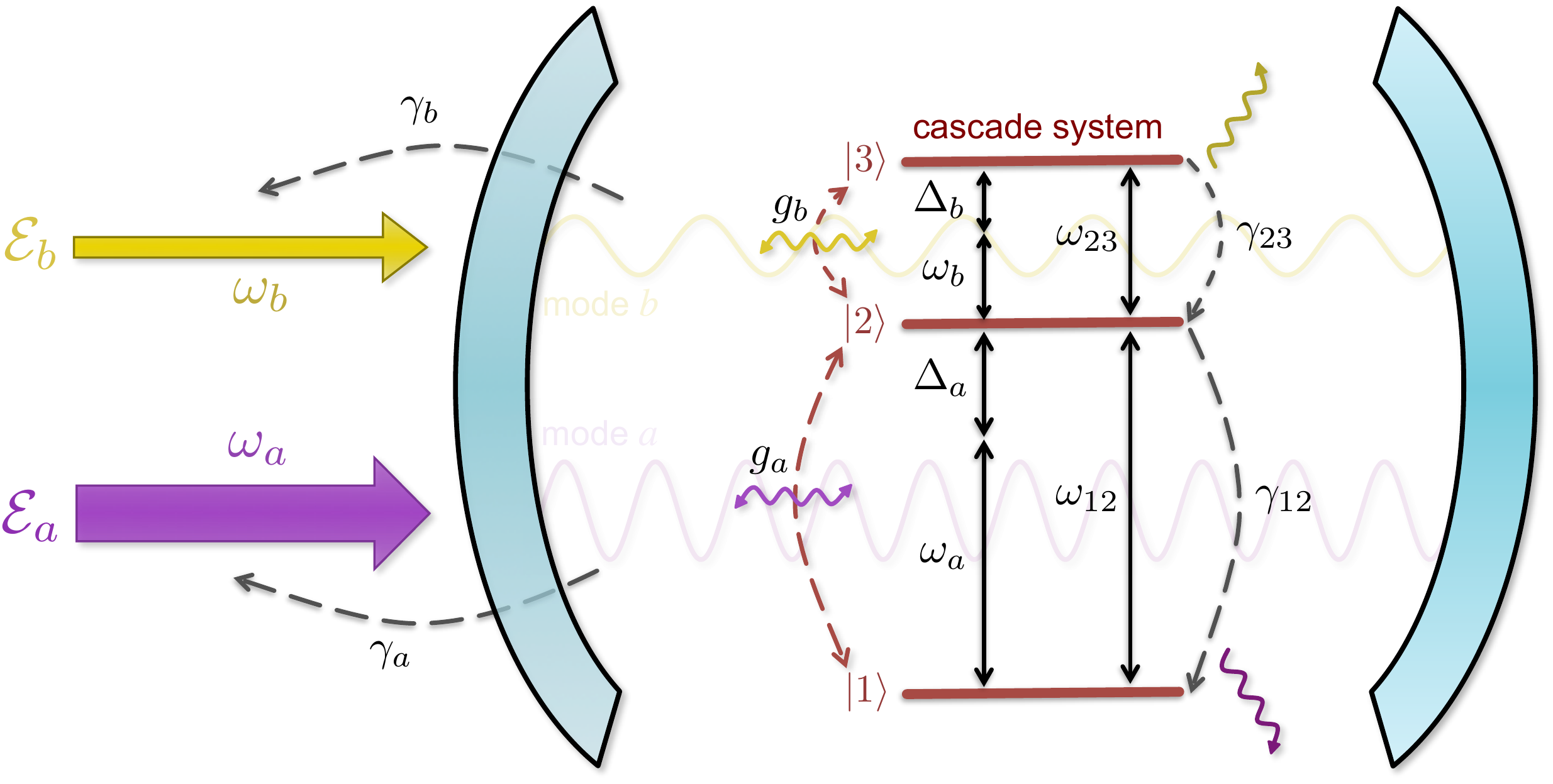}\caption{Sketch
of the system used as example: the transitions of a cascade three-level system
are coupled to two modes of a cavity, as well as to electromagnetic modes
outside the cavity which induce spontaneous emission. The cavity modes are
driven by external resonant lasers and have losses through the partially
transmitting mirror.}%
\label{FigScheme}%
\end{figure}

In order to fix ideas, let's consider one example consisting in a three-level
cascade $\Xi$ system interacting with two driven modes of a cavity which we
call $a$ and $b$, see Fig. \ref{FigScheme}. Let us first discuss the structure
of this system's Hilbert space as well as a convenient way of writing the
master equation governing its evolution, and then we will show how to code
what we need in Matlab.

\subsection{Hilbert space structure and master equation}

The complete Hilbert space of this system can be written as $\mathcal{H}%
=\mathcal{H}_{\Xi}\otimes\mathcal{H}_{a}\otimes\mathcal{H}_{b}$.
$\mathcal{H}_{\Xi}$ is the subspace of the $\Xi$ system, with basis
$\{|j\rangle\}_{j=1,2,3}$, which allows us to define the\ operators
$\hat{\sigma}_{jk}=|j\rangle\langle k|$. $\mathcal{H}_{a}$ and $\mathcal{H}%
_{b}$ are the subspaces of the cavity modes, both spanned by Fock states
$\{|n\rangle\}_{n=0,1,2,...}$, from which we define the basic annihilation
operator $\hat{a}=\sum_{n=1}^{\infty}\sqrt{n}|n-1\rangle\langle n|$ for mode
$a$, and similarly for mode $b$, whose corresponding annihilation operator we
denote by $\hat{b}$. Note that $\mathcal{H}_{a}$ and $\mathcal{H}_{b}$ are
Hilbert spaces of infinite dimension, but the computer can only deal with
finite dimension; hence, we need to truncate the Fock state bases to a certain
maximum photon number, which we will denote by $N_{a}$ and $N_{b}$ for the
corresponding modes, leading to finite-dimensional bases $\{|n\rangle
\}_{n=0,1,2,...,N_{a}}$ and $\{|n\rangle\}_{n=0,1,2,...,N_{b}}$ which
approximately span $\mathcal{H}_{a}$ and $\mathcal{H}_{b}$, respectively.

As shown in the figure, we order the states of the $\Xi$ system such that
$|3\rangle$ corresponds to the excited state, $|2\rangle$ to the middle one,
and $|1\rangle$ to the ground one, taking the energy origin in the middle
state; we name $\omega_{12}$ and $\omega_{23}$ the frequencies of the
corresponding transitions. Mode $a$ connects the $|1\rangle\rightleftharpoons
|2\rangle$ transition and has resonance frequency $\omega_{a}=\omega
_{12}-\Delta_{a}$, detuned by $\Delta_{a}$ from the transition of the $\Xi$
system. Mode $b$ connects the $|2\rangle\rightleftharpoons|3\rangle$
transition and has resonance frequency $\omega_{b}=\omega_{23}-\Delta_{b}$. We
assume that both modes are driven by resonant lasers and decay through the
partially reflecting mirror at rates $\gamma_{a}$ and $\gamma_{b}$ (the other
mirror is assumed to have perfect reflectivity, although that's not important
for this simple example). Levels $|3\rangle$ and $|2\rangle$ of the $\Xi$
system might decay through modes different than the cavity ones, what causes
them spontaneous emission to levels $|2\rangle$ and $|1\rangle$, respectively,
at rates $\gamma_{23}$ and $\gamma_{12}$. All these processes are captured by
the following master equation in the Schr\"{o}dinger picture:%
\begin{equation}
\frac{d\hat{\rho}}{dt}=-\mathrm{i}\left[  \hat{H}(t),\hat{\rho}\right]
+\gamma_{a}\mathcal{L}_{a}[\hat{\rho}]+\gamma_{b}\mathcal{L}_{b}[\hat{\rho
}]+\gamma_{12}\mathcal{L}_{\sigma_{12}}[\hat{\rho}]+\gamma_{23}\mathcal{L}%
_{\sigma_{23}}[\hat{\rho}],
\end{equation}
where $\hat{H}(t)=\hat{H}_{0}+\hat{H}_{\mathrm{coupling}}+\hat{H}%
_{\mathrm{driving}}(t)$, with%
\begin{subequations}
\begin{align}
\hat{H}_{0}  &  =\omega_{a}\hat{a}^{\dagger}\hat{a}+\omega_{b}\hat{b}%
^{\dagger}\hat{b}+\omega_{23}\hat{\sigma}_{33}-\omega_{12}\hat{\sigma}_{11},\\
\hat{H}_{\mathrm{coupling}}  &  =g_{a}(\hat{a}^{\dagger}\hat{\sigma}_{12}%
+\hat{a}\hat{\sigma}_{12}^{\dagger})+g_{b}(\hat{b}^{\dagger}\hat{\sigma}%
_{23}+\hat{b}\hat{\sigma}_{23}^{\dagger}),\\
\hat{H}_{\mathrm{driving}}(t)  &  =(e^{-\mathrm{i}\omega_{a}t}\mathcal{E}%
_{a}\hat{a}^{\dagger}+e^{\mathrm{i}\omega_{a}t}\mathcal{E}_{a}^{\ast}\hat
{a})+(e^{-\mathrm{i}\omega_{b}t}\mathcal{E}_{b}\hat{b}^{\dagger}%
+e^{\mathrm{i}\omega_{b}t}\mathcal{E}_{b}^{\ast}\hat{b}),
\end{align}
and where we have introduced the notation $\mathcal{L}_{c}[\hat{\rho}%
]=2\hat{c}\hat{\rho}\hat{c}^{\dagger}-\hat{c}^{\dagger}\hat{c}\hat{\rho}%
-\hat{\rho}\hat{c}^{\dagger}\hat{c}$, given an operator $\hat{c}$. Note that
we are not writing tensor products explicitly, and hence objects like $\hat
{a}\hat{\sigma}_{12}^{\dagger}$ must be understood as $\hat{\sigma}%
_{12}^{\dagger}\otimes\hat{a}\otimes\hat{I}$; we will stick to this economic
notation except when it can lead to a misunderstanding or we want to show the
underlaying tensor product structure of the Hilbert space explicitly for some reason.

Unfortunately, there are two properties of this master equation that make it
very difficult to deal with numerically. First, it is explicitly
time-dependent through $\hat{H}_{\mathrm{driving}}(t)$. Second, for large
driving amplitudes $\mathcal{E}_{j}$ it is to be expected that the cavity
modes will get highly populated, and we will not be able to truncate the Fock
bases to small enough $N_{a}$ and $N_{b}$. Both these problems appear
typically in many systems, and up to a point can be solved by moving to a
different picture\footnote{Recall that given the master equation
(\ref{MasterEq}), where the Hamiltonian can even be time-dependent, and a
general time-dependent unitary $\hat{U}(t)$, it is simple to prove that the
transformed state $\hat{\rho}_{U}=\hat{U}^{\dagger}\hat{\rho}\hat{U}$ evolves
according to the master equation%
\begin{equation}
\frac{d\hat{\rho}_{U}}{dt}=-\mathrm{i}\left[  \hat{H}_{U},\hat{\rho}%
_{U}\right]  +\Gamma(2\hat{J}_{U}\hat{\rho}_{U}\hat{J}_{U}^{\dagger}-\hat
{J}_{U}^{\dagger}\hat{J}_{U}\hat{\rho}_{U}-\hat{\rho}_{U}\hat{J}_{U}^{\dagger
}\hat{J}_{U}),
\end{equation}
with new Hamiltonian $\hat{H}_{U}=\hat{U}^{\dagger}\hat{H}\hat{U}%
+\mathrm{i}(d\hat{U}^{\dagger}/dt)\hat{U}$ and jump operators $\hat{J}%
_{U}=\hat{U}^{\dagger}\hat{J}\hat{U}$.}, as we will learn now. The idea
consists in doing two changes of picture (it can be done at once, but it's
more clear in two steps). First, we move to a picture rotating at the laser
frequencies (which in this example coincide with the cavity frequencies); this
is defined by the unitary transformation $\hat{U}(t)=\exp(-\mathrm{i}%
H_{\mathrm{c}}t)$ with $H_{\mathrm{c}}=\omega_{a}\hat{a}^{\dagger}\hat
{a}+\omega_{b}\hat{b}^{\dagger}\hat{b}+\omega_{b}\hat{\sigma}_{33}-\omega
_{a}\hat{\sigma}_{11}$, so that the transformed state $\hat{\rho}_{U}=\hat
{U}^{\dagger}\hat{\rho}\hat{U}$ evolves according to\footnote{Note that we use
here $d\hat{U}^{\dagger}/dt=\mathrm{i}H_{\mathrm{c}}U^{\dagger}$, and%
\begin{equation}
\hat{U}^{\dagger}\hat{a}\hat{U}=e^{-\mathrm{i}\omega_{a}t}\hat{a}\text{,
\ \ \ \ }\hat{U}^{\dagger}\hat{b}\hat{U}=e^{-\mathrm{i}\omega_{b}t}\hat
{b}\text{, \ \ \ \ }\hat{U}^{\dagger}\hat{\sigma}_{12}\hat{U}=e^{-\mathrm{i}%
\omega_{a}t}\hat{\sigma}_{12}\text{, \ \ \ \ and \ \ \ }\hat{U}^{\dagger}%
\hat{\sigma}_{23}\hat{U}=e^{-\mathrm{i}\omega_{b}t}\hat{\sigma}_{23},
\end{equation}
easy to prove from the Baker-Campbell-Haussdorf lemma%
\begin{equation}
e^{\hat{B}}\hat{A}e^{-\hat{B}}=\sum_{n=0}^{\infty}\frac{1}{n!}%
\underset{n}{\underbrace{[\hat{B},[\hat{B},...,[\hat{B},}}\hat{A}%
\underset{n}{\underbrace{]...]]}},
\end{equation}
and the commutators $[\hat{a},\hat{a}^{\dagger}]=1$ and $[\hat{\sigma}%
_{jk},\hat{\sigma}_{lm}]=\delta_{kl}\hat{\sigma}_{jm}-\delta_{mj}\hat{\sigma
}_{lk}$.}%
\end{subequations}
\begin{equation}
\frac{d\hat{\rho}_{U}}{dt}=-\mathrm{i}\left[  \hat{H}_{U},\hat{\rho}%
_{U}\right]  +\gamma_{a}\mathcal{L}_{a}[\hat{\rho}_{U}]+\gamma_{b}%
\mathcal{L}_{b}[\hat{\rho}_{U}]+\gamma_{12}\mathcal{L}_{\sigma_{12}}[\hat
{\rho}_{U}]+\gamma_{23}\mathcal{L}_{\sigma_{23}}[\hat{\rho}_{U}],
\end{equation}
with $\hat{H}_{U}=\hat{H}_{\Delta}+\hat{H}_{\mathrm{coupling}}+\hat
{H}_{\mathrm{d}}$, where%
\begin{equation}
\hat{H}_{\Delta}=\Delta_{b}\hat{\sigma}_{33}-\Delta_{a}\hat{\sigma}%
_{11},\text{ \ \ \ \ and \ \ \ \ }\hat{H}_{\mathrm{d}}=(\mathcal{E}_{a}\hat
{a}^{\dagger}+\mathcal{E}_{a}^{\ast}\hat{a})+(\mathcal{E}_{b}\hat{b}^{\dagger
}+\mathcal{E}_{b}^{\ast}\hat{b}).
\end{equation}
Hence, we see that in this picture the master equation becomes
time-independent. From this new picture we move to another one in which, in
loose terms, the photons generated by the coherent drivings are already taken
into account, so that we don't need to `count' them in the simulation. More
specifically, this picture is defined by the unitary (displacement)
transformation $\hat{D}[\alpha(t),\beta(t)]=\exp[\alpha(t)\hat{a}^{\dag
}-\alpha^{\ast}(t)\hat{a}+\beta(t)\hat{b}^{\dag}-\beta^{\ast}(t)\hat{b}]$,
which depends on two time-dependent amplitudes $\alpha(t)$ and $\beta(t)$ that
will be chosen later. In this case, the transformed state $\hat{\rho}_{D}%
=\hat{D}^{\dagger}\hat{\rho}_{U}\hat{D}$ evolves according to\footnote{This
can be proved by using%
\begin{align}
\frac{d\hat{D}^{\dagger}}{dt}  &  =(\dot{\alpha}\partial_{\alpha}+\dot{\alpha
}^{\ast}\partial_{\alpha^{\ast}}+\dot{\beta}\partial_{\beta}+\dot{\beta}%
^{\ast}\partial_{\beta^{\ast}})e^{-\alpha\alpha^{\ast}/2}e^{-\alpha\hat
{a}^{\dag}}e^{\alpha^{\ast}\hat{a}}e^{-\beta\beta^{\ast}/2}e^{-\beta\hat
{b}^{\dag}}e^{\beta^{\ast}\hat{b}}\\
&  =\left(  \dot{\alpha}^{\ast}\hat{a}-\dot{\alpha}\hat{a}^{\dag}+\dot{\beta
}^{\ast}\hat{b}-\dot{\beta}\hat{b}^{\dag}+\frac{\dot{\alpha}^{\ast}\alpha
-\dot{\alpha}\alpha^{\ast}+\dot{\beta}^{\ast}\beta-\dot{\beta}\beta^{\ast}}%
{2}\right)  \hat{D}^{\dagger},\nonumber
\end{align}
together with%
\begin{equation}
\hat{D}^{\dagger}\hat{a}\hat{D}=\hat{a}+\alpha(t)\text{ \ \ \ \ and
\ \ \ \ }\hat{D}^{\dagger}\hat{b}\hat{D}=\hat{b}+\beta(t).
\end{equation}
}%
\begin{align}
\frac{d\hat{\rho}_{D}}{dt}  &  =-\mathrm{i}\left[  \hat{H}_{D}(t),\hat{\rho
}_{D}\right]  +\gamma_{a}\mathcal{L}_{a}[\hat{\rho}_{D}]+\gamma_{b}%
\mathcal{L}_{b}[\hat{\rho}_{D}]+\gamma_{12}\mathcal{L}_{\sigma_{12}}[\hat
{\rho}_{D}]+\gamma_{23}\mathcal{L}_{\sigma_{23}}[\hat{\rho}_{D}]\\
&  +\left[  (\mathcal{E}_{a}-\gamma_{a}\alpha-\dot{\alpha})\hat{a}^{\dag
}+(\mathcal{E}_{b}-\gamma_{b}\beta-\dot{\beta})\hat{b}^{\dag}-\mathrm{H.c.}%
,\hat{\rho}_{D}\right]  ,\nonumber
\end{align}
where $\hat{H}_{D}(t)=\hat{H}_{\Delta}+\hat{H}_{\mathrm{coupling}}+\hat
{H}_{\mathrm{Rabi}}(t)$, with%
\begin{equation}
\hat{H}_{\mathrm{Rabi}}=g_{a}[\alpha^{\ast}(t)\hat{\sigma}_{12}+\alpha
(t)\hat{\sigma}_{12}^{\dagger}]+g_{b}[\beta^{\ast}(t)\hat{\sigma}_{23}%
+\beta(t)\hat{\sigma}_{23}^{\dagger}]\text{.}%
\end{equation}
This master equation suggests choosing $\alpha$ and $\beta$ such that its last
term is cancelled, that is, as solutions of $\dot{\alpha}=\mathcal{E}%
_{a}-\gamma_{a}\alpha$ and $\dot{\beta}=\mathcal{E}_{b}-\gamma_{b}\beta$:
\begin{equation}
\alpha(t)=\alpha(0)e^{-\gamma_{a}t}+\frac{\mathcal{E}_{a}}{\gamma_{a}}\left(
1-e^{-\gamma t}\right)  \text{ \ \ \ \ and \ \ \ \ }\beta(t)=\beta
(0)e^{-\gamma_{b}t}+\frac{\mathcal{E}_{b}}{\gamma_{b}}\left(  1-e^{-\gamma
_{b}t}\right)  .
\end{equation}
Note that with this change of picture we have introduced time dependence in
the master equation; however, since we are only interested in the long-time
behavior of the system (steady state), and moreover, we can choose $\alpha$
and $\beta$ at will, we can take the $t\gg\gamma_{a},\gamma_{b}$ limit in the
previous equation, in which case the displacements become time independent,
$\alpha=\mathcal{E}_{a}/\gamma_{a}$ and $\beta=\mathcal{E}_{b}/\gamma_{b}$,
and so does the master equation, which takes the final form%
\begin{equation}
\frac{d\hat{\rho}_{D}}{dt}=-\mathrm{i}\left[  \hat{H}_{D},\hat{\rho}%
_{D}\right]  +\gamma_{a}\mathcal{L}_{a}[\hat{\rho}_{D}]+\gamma_{b}%
\mathcal{L}_{b}[\hat{\rho}_{D}]+\gamma_{21}\mathcal{L}_{\sigma_{12}}[\hat
{\rho}_{D}]+\gamma_{32}\mathcal{L}_{\sigma_{23}}[\hat{\rho}_{D}],
\label{MasterEqExample}%
\end{equation}
with $\hat{H}_{D}=\hat{H}_{\Delta}+\hat{H}_{\mathrm{coupling}}+\hat
{H}_{\mathrm{Rabi}}$, being%
\begin{subequations}
\begin{align}
\hat{H}_{\Delta}  &  =\Delta_{b}\hat{\sigma}_{33}-\Delta_{a}\hat{\sigma}%
_{11},\\
\hat{H}_{\mathrm{coupling}}  &  =g_{a}(\hat{a}^{\dagger}\hat{\sigma}_{12}%
+\hat{a}\hat{\sigma}_{12}^{\dagger})+g_{b}(\hat{b}^{\dagger}\hat{\sigma}%
_{23}+\hat{b}\hat{\sigma}_{23}^{\dagger}),\\
\hat{H}_{\mathrm{Rabi}}  &  =(\Omega_{a}^{\ast}\hat{\sigma}_{12}+\Omega
_{a}\hat{\sigma}_{12}^{\dagger})+(\Omega_{b}^{\ast}\hat{\sigma}_{23}%
+\Omega_{b}\hat{\sigma}_{23}^{\dagger}),
\end{align}
where we have introduced the Rabi frequencies $\Omega_{a}=g_{a}\mathcal{E}%
_{a}/\gamma_{a}$ and $\Omega_{b}=g_{b}\mathcal{E}_{b}/\gamma_{b}$.

\subsection{Coding the problem in Matlab}

In the following we will learn how to code the previous problem in Matlab,
with the aim of finding the steady state of master equation
(\ref{MasterEqExample}), and compute certain interesting objects and
quantities derived from it.

One usually starts by defining the basic operators in the complete Hilbert
space. For this, we first need to choose the bases of the different subspaces
and order their elements; in our case, we take the bases that we introduced at
the beginning of the previous section, ordered as we did (in increasing number
of their excitation number). In particular, the basis $\{|1\rangle
,|2\rangle,|3\rangle\}$ associated to the energy levels of the $\Xi$ system
spans $\mathcal{H}_{\Xi}$, while the Fock bases $\{|0\rangle,|1\rangle
,...,|N_{a}\rangle\}$ and $\{|0\rangle,|1\rangle,...,|N_{b}\rangle\}$ span
$\mathcal{H}_{a}$ and $\mathcal{H}_{b}$, respectively. Defining the identity
of dimension 3, the representation of eigenvector $|1\rangle\in\mathcal{H}%
_{\Xi}$ corresponds to its first column, while the one of $|3\rangle
\in\mathcal{H}_{\Xi}$ to its third column. Similarly, defining the identity of
dimension $N_{a}+1$, the representation of Fock state $|0\rangle\in
\mathcal{H}_{a}$ corresponds to its first column, while that of $|N_{A}%
\rangle\in\mathcal{H}_{a}$ to its last column, and the same for mode $b$.

As for the basic operators, let's start from the ones acting on the cascade
subspace, the transition operators $\hat{\sigma}_{jk}=|j\rangle\langle k|$.
Given the vector representations $\{\mathbf{v}_{1},\mathbf{v}_{2}%
,\mathbf{v}_{3}\}$ of the basis elements $\{|1\rangle,|2\rangle,|3\rangle\}$,
the matrix representation of these operators is obtained in Matlab as
$\boldsymbol{\sigma}_{jk}=\mathbf{v}_{j}\mathbf{v}_{k}^{\dagger}$, which has a
single one at position $(j,k)$. As for the bosonic operators $\hat{a}$ and
$\hat{b}$, note that their matrix elements are $\{a_{mn}=\langle m|\hat
{a}|n\rangle=\sqrt{n}\delta_{m+1,n}\}_{n,m=0,1,...,N_{a}}$ and $\{b_{mn}%
=\langle m|\hat{b}|n\rangle=\sqrt{n}\delta_{m+1,n}\}_{n,m=0,1,...,N_{b}}$,
where in these expressions the basis vectors are Fock states in the
corresponding subspaces; hence, their matrix representations are%
\end{subequations}
\begin{equation}
\mathbf{a}=\left(
\begin{array}
[c]{ccccc}%
0 & 1 & 0 & \cdots & 0\\
0 & 0 & \sqrt{2} & \cdots & 0\\
\vdots & \vdots & \vdots & \ddots & \vdots\\
0 & 0 & 0 & \cdots & \sqrt{N_{a}}%
\end{array}
\right)  \text{ \ \ \ \ and \ \ \ }\mathbf{b}=\left(
\begin{array}
[c]{ccccc}%
0 & 1 & 0 & \cdots & 0\\
0 & 0 & \sqrt{2} & \cdots & 0\\
\vdots & \vdots & \vdots & \ddots & \vdots\\
0 & 0 & 0 & \cdots & \sqrt{N_{b}}%
\end{array}
\right)  \text{,}%
\end{equation}
with the square root of the excitation numbers in the first upper diagonal.
Note that these are the representations of the operators in their respective
subspaces, and they have to be (tensor) multiplied by the identity in the rest
of subspaces to get their representations in the complete Hilbert space, e.g.,
$\mathbf{I}^{(3)}\otimes\mathbf{a}\otimes\mathbf{I}^{(N_{b}+1)}$ in the case
of the annihilation operator of the $a$ cavity mode.

With all these considerations and the general constructions of the previous
sections, we can start writing the Matlab code. In the following, we will go
through the main parts of the code, explaining it and writing it explicitly so
that it can be copied directly to a Matlab script (a simple text file saved
with \textquotedblleft.m\textquotedblright\ extension); in any case, the whole
script can be found as part of the supplemental material.

We start by giving values to the model parameters $\Delta_{a}$, $\Delta_{b}$,
$g_{a}$, $g_{b}$, $\gamma_{12}$, $\gamma_{23}$, $\gamma_{a}$, $\gamma_{b}$,
$\Omega_{a}$, and $\Omega_{b}$:

\begin{quotation}
\texttt{Deltaa = 0; \%detuning of mode a}

\texttt{Deltab = 0; \%detuning of mode b}

\texttt{ga = 1; \%coupling mode a}

\texttt{gb = 1; \%coupling mode b}

\texttt{gamma12 = 1; \%spontaneous decay rate from 2 to 1}

\texttt{gamma23 = 1; \%spontaneous decay rate from 3 to 2}

\texttt{gammaa = 3; \%cavity damping rate of mode a}

\texttt{gammab = 3; \%cavity damping rate of mode b}

\texttt{Omegaa = 20; \%Rabi frequency driving transition 1-2}

\texttt{Omegab = 5; \%Rabi frequency driving transition 2-3}
\end{quotation}

Note that anything written after the \textquotedblleft\texttt{\%}%
\textquotedblright\ symbol (in the same line) is understood as a comment by
Matlab, and not executed. On the other hand, the semicolons \textquotedblleft%
\texttt{;}\textquotedblright\ prevent the expression from appearing in the
main command window (try removing one, and you'll see how the output of the
line is printed on screen). We have chosen simple values for the parameters,
leading to intuitive physical behavior of the system. In particular, the
cavity modes are on resonance with their corresponding transitions, the
couplings and spontaneous emission rates are on the same order, but smaller
than the damping through the mirrors, and the Rabi frequencies are the
dominant parameters, but with the lower transition driven more strongly. Under
such conditions, it is to be expected that the population of the $\Xi$ system
will be almost equally distributed between its ground and middle states, with
just a little bit in the excited state, and this is exactly what we will see later.

Let's now define the parameters related to the dimension of the Hilbert space:

\begin{quotation}
\texttt{Na = 4; \%Fock basis truncation for mode a}

\texttt{Nb = 2; \%Fock basis truncation for mode b}

\texttt{dima = Na+1; \%dimension of mode a Hilbert space}

\texttt{dimb = Nb+1; \%dimension of mode b Hilbert space}

\texttt{dims = 3; \%dimension of cascade system Hilbert space}

\texttt{dimtot = 3*dima*dimb; \%dimension of the total Hilbert space}
\end{quotation}

\begin{flushleft}
Note that one needs to check that the truncations $N_{a}$ and $N_{b}$ are
enough, by going to larger numbers, and confirming that the quantities of
interest have converged.
\end{flushleft}

Now we can start defining the matrix representations of the different
operators. We start with the identity operators in the different spaces:

\begin{quotation}
\texttt{Ia = speye(dima); \%identity on mode a subspace}

\texttt{Ib = speye(dimb); \%identity on mode b subspace}

\texttt{Is = speye(3); \%identity on cascade system subspace}

\texttt{Itot = speye(dimtot); \%identity on the complete Hilbert space}
\end{quotation}

Note that we have chosen to define them in sparse form to save memory (in full
form we would just replace \texttt{speye} by \texttt{eye}). The annihilation
operators for the cavity modes are then written in their respective subspaces as

\begin{quotation}
\texttt{a = spdiags(sqrt(0:Na)',1,dima,dima);}

\texttt{b = spdiags(sqrt(0:Nb)',1,dimb,dimb);}
\end{quotation}

\begin{flushleft}
in sparse form, or
\end{flushleft}

\begin{quotation}
\texttt{a = diag(sqrt(1:Na),1);}

\texttt{b = diag(sqrt(1:Nb),1);}
\end{quotation}

\begin{flushleft}
in full form. In the complete Hilbert space $\mathcal{H}_{\Xi}\otimes
\mathcal{H}_{a}\otimes\mathcal{H}_{b}$, these operators are coded as
\end{flushleft}

\begin{quotation}
\texttt{a = kron(Ib,kron(a,Is));}

\texttt{b = kron(b,kron(Ia,Is));}
\end{quotation}

\begin{flushleft}
as we learned in Section \ref{Composite}.
\end{flushleft}

In order to code the transition operators $\hat{\sigma}_{jk}$ of the cascade
system, it is convenient to first define the vector representation of its
basis elements, what we do as

\begin{quotation}
\texttt{v1 = Is(:,1); \%ground state of the cascade system}

\texttt{v2 = Is(:,2); \%middle state of the cascade system}

\texttt{v3 = Is(:,3); \%excited state of the cascade system}
\end{quotation}

\begin{flushleft}
Once we have the basis vectors, we can code the transition operators in the
complete Hilbert space as
\end{flushleft}

\begin{quotation}
\texttt{s11 = kron(Ib,kron(Ia,v1*v1')); \%sigma\_\{11\}}

\texttt{s22 = kron(Ib,kron(Ia,v2*v2')); \%sigma\_\{22\}}

\texttt{s33 = kron(Ib,kron(Ia,v3*v3')); \%sigma\_\{33\}}

\texttt{s12 = kron(Ib,kron(Ia,v1*v2')); \%sigma\_\{12\}}

\texttt{s13 = kron(Ib,kron(Ia,v1*v3')); \%sigma\_\{13\}}

\texttt{s23 = kron(Ib,kron(Ia,v2*v3')); \%sigma\_\{23\}}
\end{quotation}

Having the matrix representations of the fundamental operators, we are in
conditions to code the Liouvillian as a matrix in superspace. For this, it is
convenient to first code the Hamiltonian, what we do as

\begin{quotation}
\texttt{\%Build the term containing the detunings:}

\texttt{HDelta = Deltab*s33-Deltaa*s11;}

\texttt{\%the coupling terms:}

\texttt{Hcoupling = ga*(a'*s12+a*s12') + gb*(b'*s23+b*s23');}

\texttt{\%and the Rabi terms:}

\texttt{HRabi = (conj(Omegaa)*s12+Omegaa*s12') +
(conj(Omegab)*s23+Omegab*s23');}

\texttt{\%from which we build up the total Hamiltonian:}

\texttt{H = HDelta+Hcoupling+HRabi;}
\end{quotation}

\begin{flushleft}
Next we code the different dissipative pieces of the Liouvillian. As we
learned in Section \ref{MatlabImp}, this can be done as
\end{flushleft}

\begin{quotation}
\texttt{\%Damping term of mode a:}

\texttt{La =
gammaa*(2*kron(conj(a),a)-kron(Itot,a'*a)-kron(a.'*conj(a),Itot));}

\texttt{\%damping term of mode b:}

\texttt{Lb =
gammab*(2*kron(conj(b),b)-kron(Itot,b'*b)-kron(b.'*conj(b),Itot));}

\texttt{\%radiative decay of the lower transition of the cascade system:}

\texttt{L12 =
gamma12*(2*kron(conj(s12),s12)-kron(Itot,s12'*s12)-kron(s12.'*conj(s12),Itot));}%

\texttt{\%radiative decay of the upper transition of the cascade system:}

\texttt{L23 = gamma23*(2*kron(conj(s23),s23)-kron(Itot,s23'*s23)-kron(s23.}%
'\texttt{*conj(s23),Itot));}
\end{quotation}

\begin{flushleft}
Once we have the Hamiltonian and the dissipative pieces, we then build the
total Liouvillian as
\end{flushleft}

\begin{quotation}
\texttt{L = -1i*kron(Itot,H)+1i*kron(H.',Itot)+La+Lb+L12+L23; \%total
Liouvillian}
\end{quotation}

\begin{flushleft}
as given by expression (\ref{LinMatlab}).
\end{flushleft}

At this point we have managed to code the matrix representation of the
Liouvillian in superspace. Now, we proceed to evaluate its steady state in the
different ways that we introduced in Section \ref{MatlabImp} . For each
method, given the steady state which we denote here by $\hat{\rho}%
_{\mathrm{S}}$, we compute the populations $\mathrm{tr}\{\hat{\sigma}_{jj}%
\hat{\rho}_{\mathrm{S}}\}$, $\mathrm{tr}\{\hat{a}^{\dagger}\hat{a}\hat{\rho
}_{\mathrm{S}}\}$, and $\mathrm{tr}\{\hat{b}^{\dagger}\hat{b}\hat{\rho
}_{\mathrm{S}}\}$. At the end we will see that all the methods give the same populations.

As a first method we find the eigenvector with zero eigenvalue via sparse
diagonalization, as explained in Section \ref{MatlabImp}. The code looks like

\begin{quotation}
\texttt{[rhoS1,lambda0] = eigs(L,1,`LR'); \%find eigenvector with largest real
part}

\texttt{eigen0 = lambda0 \%check that the eigenvalue is 0}

\texttt{rhoS1 = reshape(rhoS1,dimtot,dimtot); \%reshape eigenvector into a
matrix}

\texttt{rhoS1 = rhoS1/trace(rhoS1); \%normalize}

\texttt{Pop1 = [trace(s11*rhoS1) trace(s22*rhoS1) trace(s33*rhoS1)...}

\ \ \ \ \ \ \ \ \ \ \ \ \ \ \ \ \texttt{trace(a'*a*rhoS1) trace(b'*b*rhoS1)];
\%evaluate populations}
\end{quotation}

\begin{flushleft}
The second line prints out the eigenvalue of the Liouvillian matrix with the
largest real part, which should appear in Matlab's command window as
\end{flushleft}

\begin{quotation}
\texttt{eigen0 =}

\texttt{ -9.6655e-15 - 7.7851e-15i}
\end{quotation}

\begin{flushleft}
Note that this is basically zero within the numerical error, just as expected.
Note also that we have introduced the three dots \textquotedblleft%
\texttt{...}\textquotedblright, which is just a way of telling Matlab that the
expression is too long, and it continues in the next line, so lines connected
by three dots are understood as a single line by Matlab.
\end{flushleft}

Let's consider now the method which uses the full diagonalization of the
Liouvillian matrix. We can code it as

\begin{quotation}
\texttt{tic \%start counting time}

\texttt{[V,D] = eig(full(L)); \%find full eigensystem of the Liouvillian}

\texttt{t\_FullDiag = toc \%time lapsed since the previous tic}

\texttt{lambdav = diag(D); \%eigenvalues}

\texttt{\%sort eigenvalues in descending order of the real part:}

\texttt{[x,y] = sort(real(lambdav),`descend'); \%y stores the permutation to
rearrange}

\texttt{V = V(:,y); \%sort the eigenvectors}

\texttt{lambdav = lambdav(y); \%sort the eigenvalues}

\texttt{eigenv = lambdav(1:5) \%show the first 5 eigenvalues}

\texttt{rhoS2 = V(:,1); \%the steady state should be the first eigenvector}

\texttt{rhoS2 = reshape(rhoS2,dimtot,dimtot); \%reshape it as a matrix}

\texttt{rhoS2 = rhoS2/trace(rhoS2); \%normalize it}

\texttt{Pop2 = [trace(s11*rhoS2) trace(s22*rhoS2) trace(s33*rhoS2)...}

\ \ \ \ \ \ \ \ \ \ \ \ \ \ \ \ \texttt{trace(a'*a*rhoS2) trace(b'*b*rhoS2)];
\%evaluate populations}
\end{quotation}

\begin{flushleft}
We have introduced the functions \texttt{tic} and \texttt{toc}, which allow to
check the time that Matlab needed to evaluate the instructions between them.
The rest just follows the recipe that we learned in Section \ref{MatlabImp}.
This piece of the code prints out the following lines in Matlab's command window:
\end{flushleft}

\begin{quotation}
\texttt{t\_FullDiag =}

\texttt{ 35.713}

\medskip\texttt{eigenv =}

\texttt{ 1.1758e-15 - 1.9065e-14i}

\texttt{ -1.0631 + 3.1308e-14i}

\texttt{ -1.5594 - 20.62i}

\texttt{ -1.5594 + 20.62i}

\texttt{ -1.5596 - 20.617i}
\end{quotation}

\begin{flushleft}
The first quantity is the time needed to perform the full diagonalization of
the Liouvillian (in seconds); you can check when running the whole code that
35 seconds is approximately 90\% of the whole time. The next quatities
correspond to the eigenvalues with the largest real part; note that only one
is zero (within the numerical error), and the rest have all negative real
parts, so we see that we really have a unique steady state. You can check that
the instruction \texttt{eigs(L,5,`LR')} gives the same 5 eigenvalues, but 60
times faster, showing the power of working with sparse matrices.
\end{flushleft}

As a final method, we code the one in which one equation defining the steady
state is substituted by the normalization condition, as explained in Section
\ref{MatlabImp}. It can be done as follows:

\begin{quotation}
\texttt{Isuper = eye(dimtot*dimtot); \%define the identity in superspace}

\texttt{l = 1; \%pick the index of the diagonal element whose equation we want
to replace}

\texttt{sl = l+(l-1)*dimtot; \%corresponding index in superspace}

\texttt{gamma = 1; \%constant by which we multiply the normalization
condition}

\medskip\texttt{L0 = full(L); \%we first define L0 as the Liouvillian in
non-sparse form}

\texttt{\%And then replace the chosen row by the part of the normalization
condition:}

\texttt{L0(sl,:) = gamma*Itot(:);}

\texttt{\%Define the vector encoding the other part of the normalization
condition:}

\texttt{w0 = gamma*Isuper(:,sl);}

\medskip\texttt{rhoS3 = L0$\backslash$w0; \%steady state in terms of the
inverse of L0}

\texttt{rhoS3 = reshape(rhoS3,dimtot,dimtot); \%reshape it as a matrix}

\texttt{tr3 = trace(rhoS3) \%check the trace, which should be 1 by
construction}

\medskip\texttt{rhoS4 = linsolve(L0,w0); \%steady state using the linear
solver of Matlab}

\texttt{rhoS4 = reshape(rhoS4,dimtot,dimtot); \%reshape it as a matrix}

\texttt{tr4 = trace(rhoS4) \%check the trace, which should be 1 by
construction}

\medskip\texttt{\%We finally evaluate the populations with both states}

\texttt{Pop3 = [trace(s11*rhoS3) trace(s22*rhoS3) trace(s33*rhoS3)...}

\ \ \ \ \ \ \ \ \ \ \ \ \ \ \ \ \texttt{trace(a'*a*rhoS3) trace(b'*b*rhoS3)];
\%populations from rhoS3}

\texttt{Pop4 = [trace(s11*rhoS4) trace(s22*rhoS4) trace(s33*rhoS4)...}

\ \ \ \ \ \ \ \ \ \ \ \ \ \ \ \ \texttt{trace(a'*a*rhoS4) trace(b'*b*rhoS4)];
\%populations from rhoS4}
\end{quotation}

\begin{flushleft}
Note that we find the steady state by solving its defining equation in the two
different ways explained in Section \ref{MatlabImp}: either by inversion of
the modified Liouvillian or using Matlab's linear solver. The code prints out
in Matlab's command window the trace of the density matrices obtained through
both methods, which should be 1 by construction. You can check that this is
indeed the case.
\end{flushleft}

Next in the code, we evaluate some reduced states as an example of how to code
the partial trace. We start from the steady state evaluated via sparse
diagonalization, rearranged as a multidimensional array as

\begin{quotation}
\texttt{rhoMDA = reshape(rhoS1,dims,dima,dimb,dims,dima,dimb);}
\end{quotation}

From this, we find the reduced state of the cavity modes by tracing out the
$\Xi$ system as

\begin{quotation}
\texttt{rhoab = permute(rhoMDA,[2,3,5,6,1,4]); \%move cascade indices to the
end}

\texttt{rhoab = reshape(rhoab,dima*dimb*dima*dimb,dims*dims); \%reshape as a
matrix}

\texttt{rhoab = rhoab*Is(:); \%trace out the cascade subspace}

\texttt{\%Reshape \texttt{the superspace vector }as a multidimensional array:}

\texttt{rhoab = reshape(rhoab,dima,dimb,dima,dimb);}

\texttt{\%and build the reduced density matrix in the a+b subspace: }

\texttt{rhoab = reshape(rhoab,dima*dimb,dima*dimb);}
\end{quotation}

We can also trace out the cavity modes, to find the reduced state of the $\Xi$ system:

\begin{quotation}
\texttt{rhos = permute(rhoMDA,[1,4,2,3,5,6]); \%move cavity indices to the
end}

\texttt{rhos = reshape(rhos,dims*dims,dima*dimb*dima*dimb); \%reshape as a
matrix}

\texttt{Iab = eye(dima*dimb); \%define identity matrix in the a+b subspace}

\texttt{rhos = rhos*Iab(:); \%trace out the cavity modes}

\texttt{\%Reshape superspace vector into the reduced matrix in the cascade
subspace:}

\texttt{rhos = reshape(rhos,dims,dims);}
\end{quotation}

Finally, we find the reduced state of each cavity mode from their combined
reduced state found before, first for mode $a$:

\begin{quotation}
\texttt{\%Reshape their combined state as a multidimensional array:}

\texttt{rhoa = reshape(rhoab,dima,dimb,dima,dimb);}

\texttt{rhoa = permute(rhoa,[1,3,2,4]); \%move indices of the b subspace to
the end}

\texttt{rhoa = reshape(rhoa,dima*dima,dimb*dimb); \%reshape as a matrix}

\texttt{rhoa = rhoa*Ib(:); \%trace out the b mode}

\texttt{\%Reshape the superspace vector into the reduced matrix in the a
subspace}

\texttt{rhoa = reshape(rhoa,dima,dima);}
\end{quotation}

\begin{flushleft}
and then for mode $b$:
\end{flushleft}

\begin{quotation}
\texttt{rhob = reshape(rhoab,dima,dimb,dima,dimb);}

\texttt{rhob = permute(rhob,[2,4,1,3]); \%move indices of the a subspace to
the end}

\texttt{rhob = reshape(rhob,dimb*dimb,dima*dima); \%reshape as a matrix}

\texttt{rhob = rhob*Ia(:); \%trace out the a mode}

\texttt{\%Reshape the superspace vector into the reduced matrix in the b
subspace:}

\texttt{rhob = reshape(rhob,dimb,dimb);}
\end{quotation}

Now that we have found the reduced steady states, let's compute the
populations from them.

\begin{quotation}
\texttt{\%Define operators in their respective subspaces:}

\texttt{ar = diag(sqrt(1:Na),1); \%annihilation operator in the a subspace}

\texttt{br = diag(sqrt(1:Nb),1); \%annihilation operator in the b subspace}

\texttt{s11r = v1*v1'; \%sigma\_\{11\}}

\texttt{s22r = v2*v2'; \%sigma\_\{22\}}

\texttt{s33r = v3*v3'; \%sigma\_\{33\}}

\texttt{PopReduced = [trace(s11r*rhos) trace(s22r*rhos) trace(s33r*rhos)...}

\ \ \ \ \ \ \ \ \ \ \ \ \ \ \ \ \ \ \ \ \ \ \ \ \ \ \ \ \ \ \texttt{trace(ar'*ar*rhoa)
trace(br'*br*rhob)]; \%populations}
\end{quotation}

Then, we build a matrix containing the populations from all the methods as
columns (note that we take the real parts, so that the imaginary parts are not
printed out to save space on screen, but check yourself that the latter are
zero as they should be):

\begin{quotation}
\texttt{Pop = real([Pop1; Pop2; Pop3; Pop4; PopReduced]')}
\end{quotation}

\begin{flushleft}
which printed in Matlab's command window reads:
\end{flushleft}

\begin{quotation}
\texttt{Pop =}

$\ \ \ \ \ \ \ \ \
\begin{array}
[c]{rrrrr}%
\mathtt{0.45882} & \mathtt{0.45882} & \mathtt{0.45882} & \mathtt{0.45882} &
\mathtt{0.45882}\\
\mathtt{0.48438} & \mathtt{0.48438} & \mathtt{0.48438} & \mathtt{0.48438} &
\mathtt{0.48438}\\
\mathtt{0.056796} & \mathtt{0.056796} & \mathtt{0.056796} & \mathtt{0.056796}
& \mathtt{0.056796}\\
\mathtt{0.019165} & \mathtt{0.019165} & \mathtt{0.019165} & \mathtt{0.019165}
& \mathtt{0.019165}\\
\mathtt{0.0012705} & \mathtt{0.0012705} & \mathtt{0.0012705} &
\mathtt{0.0012705} & \mathtt{0.0012705}%
\end{array}
$
\end{quotation}

\begin{flushleft}
showing that all the steady states give exactly the same populations. Note
that the populations of the $\Xi$ system are what we were expecting from the
system parameters. On the other hand, note also that we have computed is not
the true cavity populations, since our state is not in the Schr\"{o}dinger
picture, but in a displaced picture where the external driving is subtracted.
Taking into account that the steady state is $\hat{U}\hat{D}\hat{\rho
}_{\mathrm{S}}\hat{D}^{\dagger}\hat{U}^{\dagger}$ in the Schr\"{o}dinger
picture, we can get the true cavity populations as
\begin{subequations}
\begin{align}
\mathrm{tr}\{\hat{a}^{\dagger}\hat{a}\hat{U}\hat{D}\hat{\rho}_{\mathrm{S}}%
\hat{D}^{\dagger}\hat{U}^{\dagger}\}  &  =\mathrm{tr}\{\hat{D}^{\dagger}%
\hat{U}^{\dagger}\hat{a}^{\dagger}\hat{a}\hat{U}\hat{D}\hat{\rho}_{\mathrm{S}%
}\}=\mathrm{tr}\{(\hat{a}^{\dagger}+\alpha^{\ast})(\hat{a}+\alpha)\hat{\rho
}_{\mathrm{S}}\}=|\alpha|^{2}+\mathrm{tr}\{\hat{a}^{\dagger}\hat{a}\hat{\rho
}_{\mathrm{S}}\}+2\operatorname{Re}\{\alpha^{\ast}\mathrm{tr}\{\hat{a}%
\hat{\rho}_{\mathrm{S}}\}\},\\
\mathrm{tr}\{\hat{b}^{\dagger}\hat{b}\hat{U}\hat{D}\hat{\rho}_{\mathrm{S}}%
\hat{D}^{\dagger}\hat{U}^{\dagger}\}  &  =\mathrm{tr}\{\hat{D}^{\dagger}%
\hat{U}^{\dagger}\hat{b}^{\dagger}\hat{b}\hat{U}\hat{D}\hat{\rho}_{\mathrm{S}%
}\}=\mathrm{tr}\{(\hat{b}^{\dagger}+\alpha^{\ast})(\hat{b}+\alpha)\hat{\rho
}_{\mathrm{S}}\}=|\alpha|^{2}+\mathrm{tr}\{\hat{b}^{\dagger}\hat{b}\hat{\rho
}_{\mathrm{S}}\}+2\operatorname{Re}\{\alpha^{\ast}\mathrm{tr}\{\hat{b}%
\hat{\rho}_{\mathrm{S}}\}\},
\end{align}
which we compute in Matlab as
\end{subequations}
\end{flushleft}

\begin{quotation}
\texttt{alpha = Omegaa/ga;}

\texttt{beta = Omegab/gb;}

\texttt{Popa = Pop1(4)+conj(alpha)*alpha+2*real(conj(alpha)*trace(a*rhoS1))}

\texttt{Popb = Pop1(5)+conj(beta)*beta+2*real(conj(beta)*trace(b*rhoS1))}
\end{quotation}

\begin{flushleft}
printing out the following result in Matlab's command window:
\end{flushleft}

\begin{quotation}
\texttt{Popa =}

\texttt{ 399.66 + 3.9078e-16i}

\medskip\texttt{Popb =}

\texttt{ 24.961 + 2.1115e-16i}
\end{quotation}

\begin{flushleft}
Hence, we see that with such strong drivings, the $\Xi$ system doesn't change
too much the cavity populations from their values expected in the absence of
coupling, $|\alpha|^{2}=400$ for mode $a$ and $|\beta|^{2}=25$ for mode $b$.
\end{flushleft}

Finally in the code, we proceed to check the entanglement between various
bipartitions of the complete system, what will give us a perfect excuse to
compute some partial transpositions. Given the state $\hat{\rho}$ of a system
whose Hilbert space we divide in two as $\mathcal{H}_{A}\otimes\mathcal{H}%
_{B}$, a necessary condition for it to be separable with respect to that
bipartition is that the partial transpose $\hat{\rho}^{T_{A}}$ is
semi-positive definite, that is, it has only positive or zero eigenvalues.
Given the eigenvalues $\{\tilde{\lambda}_{n}\}_{n}$ of $\hat{\rho}^{T_{A}}$,
we can evaluate the level of violation of such condition via the logarithmic
negativity $E_{\mathrm{LN}}=\log[1+\sum_{n}(|\tilde{\lambda}_{n}%
|-\tilde{\lambda}_{n})]$, which is one of the most common entanglement
measures available for mixed states. In the following we evaluate this
quantity for various bipartitions of our system.

Let's start with the entanglement between the $\Xi$ system and the cavity
modes. We can find the corresponding logarithmic negativity as

\begin{quotation}
\texttt{\%Given the full state as a multidimensional array,}

\texttt{\%we first transpose the cascade subspace:}

\texttt{rhoT = permute(rhoMDA,[4,2,3,1,5,6]);}

\texttt{rhoT = reshape(rhoT,dimtot,dimtot); \%reshape it as a matrix}

\texttt{Teigenv = eig(rhoT); \%compute its eigenvalues}

\texttt{logNeg = log(1+sum(abs(Teigenv)-Teigenv)); \%compute the log
negativity}
\end{quotation}

Let's compute now the entanglement between the cavity modes, what we do as

\begin{quotation}
\texttt{\%First reshape the reduced state of the cavity modes}

\texttt{\%as a multidimensional array:}

\texttt{rhoabT = reshape(rhoab,dima,dimb,dima,dimb);}

\texttt{rhoabT = permute(rhoabT,[3,2,1,4]); \%transpose the a mode subspace}

\texttt{rhoabT = reshape(rhoabT,dima*dimb,dima*dimb); \%reshape it as a
matrix}

\texttt{abTeigenv = eig(rhoabT); \%evaluate its eigenvalues}

\texttt{logNegab = log(1+sum(abs(abTeigenv)-abTeigenv)); \%compute the log
negativity}
\end{quotation}

Finally we check the entanglement between the $\Xi$ system and each of the
cavity modes individually. We start with the $a$ mode as

\begin{quotation}
\texttt{\%First we need the reduced state of the cascade system and the a
mode.}

\texttt{\%Starting from the complete state as a multidimensional array,}

\texttt{\%we move the \ indices of the b mode to the end:}

\texttt{rhoas = permute(rhoMDA,[1,2,4,5,3,6]);}

\texttt{rhoas = reshape(rhoas,dima*dims*dima*dims,dimb*dimb); \%reshape as a
matrix}

\texttt{rhoas = rhoas*Ib(:); \%trace out the b subspace}

\texttt{\%and reshape the superspace vector as a multidimensional array:}

\texttt{rhoas = reshape(rhoas,dims,dima,dims,dima); }

\texttt{rhoasT = permute(rhoas,[3,2,1,4]); \%transpose cascade subspace}

\texttt{rhoasT = reshape(rhoasT,dima*dims,dima*dims); \%reshape it as a
matrix}

\texttt{asTeigenv = eig(rhoasT); \%find eigenvalues}

\texttt{logNegas = log(1+sum(abs(asTeigenv)-asTeigenv)); \%compute the log
negativity}
\end{quotation}

\begin{flushleft}
and then for the $b$ mode as
\end{flushleft}

\begin{quotation}
\texttt{\%First we need the reduced state of the cascade system and the b
mode.}

\texttt{\%Starting from the complete state as a multidimensional array,}

\texttt{\%we move the \ indices of the a mode to the end:}

\texttt{rhobs = permute(rhoMDA,[1,3,4,6,2,5]);}

\texttt{rhobs = reshape(rhobs,dimb*dims*dimb*dims,dima*dima); \%reshape as a
matrix}

\texttt{rhobs = rhobs*Ia(:); \%trace out the a subspace}

\texttt{\%and reshape the superspace vector as a multidimensional array:}

\texttt{rhobs = reshape(rhobs,dims,dimb,dims,dimb); }

\texttt{rhobsT = permute(rhobs,[3,2,1,4]); \%transpose cascade subspace}

\texttt{rhobsT = reshape(rhobsT,dimb*dims,dimb*dims); \%reshape it as a
matrix}

\texttt{bsTeigenv = eig(rhobsT); \%find eigenvalues}

\texttt{logNegbs = log(1+sum(abs(bsTeigenv)-bsTeigenv)); \%compute the log
negativity}
\end{quotation}

Collecting all the logarithmic negativities in a single vector as

\begin{quotation}
\texttt{LogNegativities = [logNeg; logNegab; logNegas; logNegbs]}
\end{quotation}

\begin{flushleft}
we get the following printed out in Matlab's command window:
\end{flushleft}

\begin{quotation}
\texttt{LogNegativities =}

\texttt{ 0.0025892 - 1.061e-16i}

\texttt{ 2.027e-07 - 9.707e-17i}

\texttt{ 0.0017957 - 1.2758e-16i}

\texttt{ 9.2002e-05 - 1.5018e-17i}
\end{quotation}

\begin{flushleft}
This shows that there is indeed entanglement between all the bipartitions,
although it is not very big in any case (take a bell state as an example,
which has logarithmic negativity equal to $\log2\approx0.7$), consistent with
the fact that the cavity populations are not very much affected by the
coupling to the $\Xi$ system. Note in particular that the largest entanglement
is between the $\Xi$ system and the cavity modes, while there is almost no
entanglement between the cavity modes themselves. On the other hand, the
entanglement of the $\Xi$ system with the $a$ mode is much larger than that
with the $b$ mode.
\end{flushleft}

\end{document}